\documentclass[english,12pt,aps,prd,a4paper,preprintnumbers,floatfix,nofootinbib,showpacs,superscriptaddress]{revtex4-1}
\pdfoutput=1

\usepackage[english]{babel}
\usepackage{amsmath}
\usepackage{graphicx}
\usepackage{color}
\usepackage{mathrsfs}   %\mathscr{L}
\usepackage{amssymb}
\usepackage{hyperref}
\usepackage[normalem]{ulem} % for striking out text
\usepackage{dcolumn}% Align table columns on decimal point
\usepackage{bm}% bold math
\usepackage[sort&compress]{natbib}

\usepackage{latexsym}
\usepackage{amsthm}
\usepackage{amsmath}
\usepackage{wrapfig}
\usepackage{amssymb}
\usepackage{color}
\usepackage{cancel}
\usepackage{hyperref}                %lianjie
\usepackage{multirow}
\usepackage{pst-3d}                  % PSTricks 3D framed boxes
\usepackage{pst-grad}                % PSTricks gradient mode
\usepackage{pst-node}                % PSTricks nodes
\usepackage{pst-slpe}                % Improved PSTricks gradients
\usepackage{pst-plot}                %
\usepackage{pst-coil}                %
\usepackage{pst-math}                %
\usepackage{pstricks-add}            %
\usepackage{rotating}
\usepackage{geometry}
\usepackage{hhline}
\usepackage{amsmath}
\usepackage{float}
\usepackage{ulem}
\usepackage{fancybox}

\allowdisplaybreaks
\allowdisplaybreaks[1]
\allowdisplaybreaks[2]

\def\simgt{\mathrel{\lower2.5pt\vbox{\lineskip=0pt\baselineskip=0pt
           \hbox{$>$}\hbox{$\sim$}}}}
\def\simlt{\mathrel{\lower2.5pt\vbox{\lineskip=0pt\baselineskip=0pt
           \hbox{$<$}\hbox{$\sim$}}}}

\newcommand{\be}{\begin{eqnarray}}
\newcommand{\ee}{\end{eqnarray}}

\def\bea{\begin{eqnarray}}
\def\eea{\end{eqnarray}}

\newcommand{\AddrAHEP}{%
  AHEP Group, Institut de F\'{i}sica Corpuscular --
  CSIC/Universitat de Val\`{e}ncia, Parc Cient\'ific de Paterna.\\
 C/ Catedr\'atico Jos\'e Beltr\'an, 2 E-46980 Paterna (Valencia) - SPAIN}

\begin{document}

\begin{flushright}
\ \hfill\mbox{\small USTC-ICTS-17-07}\\[4mm]
\begin{minipage}{0.2\linewidth}
\normalsize
\end{minipage}
\end{flushright}

\title{ Neutrino Predictions from Generalized \\ CP
  Symmetries of Charged Leptons }

\author{Peng Chen}\email{pche@mail.ustc.edu.cn}
\affiliation{College of Information Science and Engineering,Ocean University of China, Qingdao 266100, China}
\author{Salvador Centelles Chuli\'{a}}\email{salcen@ific.uv.es}
\affiliation{\AddrAHEP}
\author{Gui-Jun Ding}\email{dinggj@ustc.edu.cn}
\affiliation{Interdisciplinary Center for Theoretical Study and Department of Modern Physics, \\
University of Science and Technology of China, Hefei, Anhui 230026, China}
\author{Rahul Srivastava}\email{rahulsri@ific.uv.es}
\affiliation{\AddrAHEP}
\author{Jos\'{e} W. F. Valle}\email{valle@ific.uv.es}
\affiliation{\AddrAHEP}

\begin{abstract}
   \vspace{1cm}

We study the implications of generalized CP transformations acting
on the mass matrices of charged leptons in a model--independent
way.  Generalized $e-\mu$, $e-\tau$ and $\mu-\tau$ symmetries are
considered in detail. In all cases the physical parameters of the
lepton mixing matrix, three mixing angles and three CP phases can
be expressed in terms of a restricted set of independent ``theory
parameters'' that characterize a given choice of CP transformation.
This leads to implications for neutrino oscillations as well as
neutrinoless double beta decay experiments.

\end{abstract}


\maketitle

%%%%%%%%%%%%%%%%%%%%%%%%%%%%%%%%%%%%%%%%%%%%%%%%%%%%%%%%%%%%%%%%%%%%%%%

\section{Introduction}

%%%%%%%%%%%%%%%%%%%%%%%%%%%%%%%%%%%%%%%%%%%%%%%%%%%%%%%%%%%%%%%%%%%%%%%

Following the discovery of neutrino
oscillations~\cite{Kajita:2016cak,McDonald:2016ixn}, particle physics
has entered probably its most exciting phase in many decades. The
observed pattern of the lepton mixing angles, at odds with those of
the quark sector, has added a further challenge to the so-called
flavour problem, a long-standing mystery in particle physics.
Over the last two decades, neutrino oscillation experiments have made
considerable progress in the determination of the neutrino oscillation
parameters, four of which are rather well measured. These include the
solar and atmospheric mass splittings, as well as two of the lepton
mixing angles $\theta_{12}$ and $\theta_{13}$~\cite{deSalas:2017kay}.
So far the octant of the angle $\theta_{23}$ is not yet well
determined, nor the phase parameters characterizing CP violation.
Concerning CP violation there are three physical phases in the
simplest unitary mixing matrix, one Dirac and the two Majorana
phases~\cite{Schechter:1980gr,Schechter:1981gk}.
Although latest results of T2K~\cite{Abe:2017uxa} and
NO$\nu$A~\cite{Adamson:2017gxd} provide a first positive hint for CP
violation, characterized by the Dirac CP phase $\delta_{CP}$ around
$3\pi/2$, its value is not yet measured with high
significance~\cite{deSalas:2017kay}.
Finally, negative searches for neutrinoless double beta decay do not
allow us to decide whether neutrinos are Majorana or Dirac particles
nor determine the two associated Majorana phases.

Understanding the pattern of neutrino mixing and CP violation
constitutes a fundamental problem in particle physics. Flavor
symmetries provide an attractive framework for explaining the
leptonic mixing angles and phases. In particular, non-Abelian
discrete flavor symmetries have been widely studied in the
literature see, for example,~\cite{babu:2002dz,Altarelli:2010gt,Ishimori:2010au,Morisi:2012fg,Morisi:2013qna,King:2013eh,King:2014nza,King:2015aea}.

It has been noted in recent years that discrete flavor symmetries can be extended so as to include a CP symmetry~\cite{Feruglio:2012cw,Holthausen:2012dk,Chen:2014tpa,Ding:2013hpa,Chen:2014wxa,Everett:2015oka,Chen:2015nha,Chen:2015siy,Chen:2016ica}.
The possible lepton mixing patterns which can be obtained from the
breaking of flavor and CP symmetry have been widely explored (see,
e.g.,~\cite{Ding:2013bpa,Hagedorn:2014wha,Ding:2014ora,Li:2015jxa,DiIura:2015kfa,Ballett:2015wia,Ding:2015rwa,Li:2016ppt,Yao:2016zev,Ivanov:2017bdx}
and references quoted therein). It is remarkable that the observed
patterns of quark and lepton flavor mixings can be simultaneously
understood from a flavor group and CP symmetry~\cite{Li:2017abz}. In
this approach the mixing is determined in terms of a few free
parameters, so that certain sum rules relate the mixing angles and CP
violation phases. These sum rules are sensitive probes to test the
discrete flavor symmetry approach in current and future neutrino
experiments~\cite{Ballett:2013wya,Petcov:2014laa,Girardi:2014faa}.

The imposition of a CP symmetry may allow us to obtain predictions
for CP violating phases. A simple and interesting example is the so
called $\mu-\tau$ reflection symmetry which exchanges a muon (tau)
neutrino with a tau (muon) antineutrino. If the neutrino mass matrix
is invariant under the action of $\mu-\tau$ reflection in the
charged lepton diagonal basis, both atmospheric mixing angle
$\theta_{23}$ and Dirac CP phase $\delta_{CP}$ are maximal and the
Majorana phases are trivial~\cite{Harrison:2002et,Grimus:2003yn,Harrison:2004he}. The
above prediction for maximal $\theta_{23}$ may be at variance with
current experiments~\cite{Abe:2017uxa,Adamson:2017gxd}.
A generalized $\mu-\tau$ reflection in the charged lepton diagonal
mass basis is suggested in~\cite{Chen:2015siy}. This can accommodate
the observed non-maximal $\theta_{23}$ together with maximal
$\delta_{CP}$. Moreover, all possible CP transformations can be
classified according to the number of zero entries~\cite{Chen:2016ica}.

In this work we shall study three kinds of simple and attractive CP
symmetries: generalized $e-\mu$, generalized $e-\tau$ and
generalized $\mu-\tau$ symmetries in the neutrino mass diagonal
basis. The rest of this paper is organized as follows: in
section~\ref{sec:symm_para} the symmetric parametrization of the
lepton mixing matrix is reviewed. In section~\ref{sec:general} we
demonstrate how to extract the lepton mixing matrix from an imposed
CP transformation, explaining the difference between Dirac and
Majorana neutrino cases. In section~\ref{sec:main_cont} we perform a
detailed study of the generalized $e-\mu$, $e-\tau$ and $\mu-\tau$
symmetries acting on the charged lepton fields. The predictions for
lepton mixing parameters are discussed, and the correlations among
mixing angles and CP phases are analyzed. In section~\ref{sec:mee}
we discuss the phenomenological implications of our scheme for
neutrinoless double beta decay. Further summary and conclusions are
given in section~\ref{sec:conclusion}.

%%%%%%%%%%%%%%%%%%%%%%%%%%%%%%%%%%%%%%%%%%%%%%%%%%%%%%%%%%%%%%%%%%%%%%

\section{\label{sec:symm_para}Lepton mixing matrix parametrization}

%%%%%%%%%%%%%%%%%%%%%%%%%%%%%%%%%%%%%%%%%%%%%%%%%%%%%%%%%%%%%%%%%%%%%%

Throughout this paper we will use the so-called ``symmetric
parametrization'' of the lepton mixing matrix. In this
parametrization~\cite{Schechter:1980gr}, the lepton mixing matrix $U$
can be written as~\cite{Rodejohann:2011vc}~\footnote{ This is based on
  Okubo's parametrization of unitary groups and is specially convenient
  to describe both quark and lepton mixing matrices in full
  generality~\cite{Schechter:1980gr}.}
\begin{eqnarray}
U = P_0 U_{23} U_{13} U_{12}~,
\label{sympar}
\end{eqnarray}
where
$P_0 = e^{\textnormal{diag}(i \delta_1, i \delta_2, i \delta_3)}$ and
the $U_{ij}$ are the complex rotation matrices in the
$ij$-axis. For example:
\begin{eqnarray}
U_{13} = \left(\begin{array}{ccc}
\cos \theta_{13} 					&0				&	 e^{-i \phi_{13}} \sin \theta_{13} \\
0 					& 1				 					& 0 \\
-e^{i \phi_{13}} \sin \theta_{13} 	&0					&  \cos \theta_{13} 	
\end{array}\right)\
\end{eqnarray}
The mixing angles $\theta_{12}$, $\theta_{23}$ and $\theta_{13}$ can
be limited in the range $0\leq\theta_{ij}\leq\pi/2$, the CP violation
phases $\phi_{12}$ and $\phi_{13}$ can take values in the range
$0\leq\phi_{12}, \phi_{13}<\pi$, and the phase $\phi_{23}$ can take
values in the range $0\leq\phi_{23}<2\pi$. Moreover, the phases
$\delta_{1}$, $\delta_{2}$ and $\delta_{3}$ can be rotated away by
redefinitions of the left-handed charged leptons and are therefore
unphysical.  In this parametrization, the invariant describing CP
violation in conventional neutrino oscillations, takes the
form~\cite{Jarlskog:1985ht}
\begin{equation}
J_{CP}=\Im\left(U_{11}U_{33}U^{*}_{13}U^{*}_{31}\right)=\frac{1}{8}\sin2\theta_{12}\sin2\theta_{13}\sin2\theta_{23}\cos\theta_{13}\sin\left(\phi_{13}-\phi_{12}-\phi_{23}\right)\,.\\
\end{equation}
Hence the usual Dirac CP phase relevant to neutrino oscillation is simply given by
\begin{eqnarray}
\delta_{CP}=\phi_{13}-\phi_{12}-\phi_{23}\,.
\end{eqnarray}
The other two rephasing invariants associated with the Majorana
phases are~\cite{Branco:1986gr,Jenkins:2007ip,Branco:2011zb}
\begin{equation}
I_1=\Im\left(U^2_{12}U^{*2}_{11}\right),\qquad I_2=\Im\left(U^2_{13}U^{*2}_{11}\right)\,,
\end{equation}
which take the following form in the symmetric parametrization
\begin{equation}
I_{1}=-\frac{1}{4} \sin^{2} 2\theta_{12} \cos^{4} \theta_{13}\sin(2\phi_{12})\,, \quad  I_{2}=-\frac{1}{4} \sin^{2} 2 \theta_{13} \cos^{2} \theta_{12}
   \sin(2\phi_{13})\,.
\end{equation}
On the other hand, the relevant parameter characterizing the
neutrinoless double beta decay amplitude, i.e. the ``effective
Majorana mass'' parameter, $m_{ee}$, is given as
\begin{equation}
m_{ee}=\left|m_1 c_{12}^2 c_{13}^2 + m_2 c_{13}^2 s_{12}^2 e^{-2 i \phi_{12}}+ m_3 s_{13}^2 e^{-2 i \phi_{13}}\right|\,. \label{eq:mee}
\end{equation}
Notice that only the phases $\phi_{12}$ and $\phi_{13}$, but not
$\phi_{23}$, appear in $m_{ee}$.

%%%%%%%%%%%%%%%%%%%%%%%%%%%%%%%%%%%%%%%%%%%%%%%%%%%%%%%%%%%%%%%%%%%%%%

\section{\label{sec:general}General discussion}

%%%%%%%%%%%%%%%%%%%%%%%%%%%%%%%%%%%%%%%%%%%%%%%%%%%%%%%%%%%%%%%%%%%%%%

In this section we begin with a general discussion of the generalized
CP transformations, highlighting key concepts as well as setting up
our notation and conventions. Following
Refs.~\cite{Chen:2014wxa,Chen:2015siy,Chen:2016ica} we start by
defining the generalized remnant CP transformations for each fermionic
field as follows:
\begin{eqnarray}
 \psi \stackrel{CP}{\longmapsto} i X_\psi \gamma ^0 \mathcal{C} \bar{\psi} ^T, \hspace{2mm} \psi \in \{\nu_L, \nu_R, l_L, l_R \}\,.
\end{eqnarray}
Such generalized CP transformations acting on the chiral fermions will
be a symmetry of the mass term in the Lagrangian provided they satisfy
the following conditions~\footnote{Even though the X-matrix is symmetric,
  we prefer to use $X^\dagger$ instead of $X^*$ when dealing with Dirac neutrinos.},
\begin{eqnarray}
X_\psi^Tm_\psi X_\psi & = & m_\psi^*,\quad \textnormal{ for Majorana fields}\,, \label{eq:cpmaj}   \\
X_\psi^\dagger M^2_\psi X_\psi & = & M_\psi^{2 *},\quad \textnormal{ for Dirac fields, where } M^2_\psi \equiv m_\psi^\dagger \, m_\psi \,, \label{eq:cpdir}
\end{eqnarray}
where $m_{\psi}$ is written in a basis with left-handed (right-handed)
fields on the right-hand (left-hand) side. Note that the mass matrices
$m_{\psi}$ and $M^2_{\psi}$ can be diagonalized by a unitary
transformation $U_{\psi}$,
\begin{eqnarray}
\label{eq:diag_maj}&&U^T_{\psi} m_{\psi} \,U_{\psi}=\text{diag}(m_1,m_2,m_3),\quad \textnormal{ for Majorana fields}\,,\\
\label{eq:diag_dir}&&U^\dagger_{\psi} M^2_{\psi} U_{\psi}=\text{diag}(m^2_1,m^2_2,m_3^2),\quad \textnormal{for Dirac fields}\,,
\end{eqnarray}
with $m_1\neq m_2\neq m_3$. From Eq.~\eqref{eq:cpmaj},
Eq.~\eqref{eq:cpdir}, Eq.~\eqref{eq:diag_maj} and
Eq.~\eqref{eq:diag_dir}, after straightforward algebra, we find that
the unitary transformation $U_{\psi}$ is subject to the following
constraint from the imposed CP symmetry $X_{\psi}$,
\begin{equation}
\label{eq:XCons}U^\dagger_{\psi} X_{\psi} U^*_{\psi}\equiv P=\left\{\begin{array}{cc}
\text{diag}(\pm 1,\pm 1,\pm 1),& \textnormal{ for Majorana fields},\\[0.1in]
\text{diag}(e^{i\delta_e}, e^{i\delta_\mu}, e^{i\delta_\tau}),& \textnormal{for Dirac fields}\,,
\end{array}
\right.
\end{equation}
where $\delta_{e}$, $\delta_{\mu}$ and $\delta_{\tau}$ are arbitrary
real parameters~\footnote{If neutrinos are Majorana particles and the
  lightest one is massless (this possibility is still allowed by
  current experimental data) one ``$\pm$'' entry would be a complex
  phase.}. Because $X_{\psi}$ is a symmetric matrix, one can use
Takagi decomposition (note that this decomposition is not unique) to
express $X_{\psi}$ as
\begin{eqnarray}
\label{eq:Td}X_{\psi}=\Sigma \cdot \Sigma^T\,.
\end{eqnarray}
Inserting Eq.~\eqref{eq:Td} into Eq.~\eqref{eq:XCons}, we find that
the combination $P^{-\frac{1}{2}} U^{\dagger}_{\psi}\Sigma$ is a real
orthogonal matrix, i.e.,
\begin{eqnarray}
\label{eq:ugen}P^{-\frac{1}{2}}U^{\dagger}_{\psi}\Sigma\equiv O_3\,,
\end{eqnarray}
which implies
\begin{equation}
\label{ugenn}U_{\psi}=\Sigma %\sout{O_3^\dagger}
O_3^T P^{-\frac{1}{2}}\,,
\end{equation}
where $O_3$ is a generic $3\times 3$ real orthogonal matrix. As we will discuss in detail in Section \ref{sec:main_cont}, the matrix $\Sigma$ can be expressed in terms of three independent ``CP labels'', namely two of the three phases $\alpha, \beta, \gamma$ and the CP angle $\Theta$. These CP labels characterize a given generalized CP transformation. The resulting explicit form of $\Sigma$ will depend on the generalized CP symmetry under
consideration, and will be discussed in detail in Section~\ref{sec:main_cont}.
Throughout this paper we will parameterize the orthogonal matrix $O_3$ as
\begin{equation}
\label{eq:O3}
O_3  = \left(\begin{array}{ccc}
1 & 0 & 0 \\
0 & \cos\theta_1   &   \sin\theta_1 \\
0 & -\sin\theta_1  &   \cos\theta_1
\end{array}\right)
\left(\begin{array}{ccc}
\cos\theta_2   &   0    &    \sin\theta_2 \\
0   &   1   &   0 \\
-\sin\theta_2   &   0   &    \cos\theta_2
\end{array}\right)
\left(\begin{array}{ccc}
\cos\theta_3     &    \sin\theta_3    &    0 \\
-\sin\theta_3    &    \cos\theta_3    & 0   \\
0    &    0     &    1
\end{array}\right)\,.
\end{equation}
Notice that the matrix $O_3$ has the following properties
\begin{eqnarray}
\nonumber&&O_3(\theta_1+\pi,\theta_2,\theta_3)=\text{diag}(1, -1, -1)\,O_3(\theta_1,\theta_2, \theta_3)\,,\\
 \nonumber & & O_3(\theta_1, \theta_2+\pi, \theta_3) = \text{diag} ( -1, -1, 1)\, O_3(\pi- \theta_1, \theta_2,\theta_3)\,,\\
&&\label{eq:O33_char}O_3(\theta_1, \theta_2, \theta_3+\pi) = \text{diag}( 1, -1, -1)\, O_3(\theta_1, \pi-\theta_2, \theta_3)\,,
\end{eqnarray}
where the diagonal matrices can be absorbed into the matrix $P$. As a
consequence, the range of variation of the free parameters
$\theta_{1, 2, 3}$ describing a given theory, can be taken to be
$[0,\pi)$.
At this point we would like to remark that, the generalized CP
symmetries do not impose any constraint on the fermion masses.
These can always be chosen to match the required experimental
values. The predictive power of generalized CP symmetries lies in
the mixing matrix elements and their phases.

Before going to particular cases of generalized CP transformations and
their implications we would like to briefly comment about the
differences one should expect in case neutrinos are Dirac or Majorana
particles.
In previous works on generalized CP symmetries
\cite{Chen:2014wxa,Chen:2015siy,Chen:2016ica}, the neutrinos where
assumed to be Majorana particles but that may not be the case in
nature.
We now comment on the differences that arise if neutrinos are Dirac
particles.

As clear from our previous discussion, and in particular from
Eq.~\eqref{ugenn}, the only difference between Majorana and Dirac
neutrino mixing matrices (for a given CP symmetry) is in the diagonal
unitary matrix $P$ that appears in the right side of the lepton
mixing matrix.
For the case of Majorana fields, $P$ is real, while in the Dirac case
it is a general diagonal matrix of phases. Using the technique described in
\cite{Schechter:1980gr}, it is easy to show that $\delta_{CP}$ of a
given unitary matrix $U$ is the same as $\delta_{CP}$ associated to
another unitary matrix given by $U \cdot P$, where $P$ is a diagonal
matrix of phases. This implies that for any choice of a given
generalized CP symmetry:
\begin{itemize}

\item All the mixing parameters characterizing neutrino oscillations,
  i.e., $\theta_{12}$, $\theta_{13}$, $\theta_{23}$ and $\delta_{CP}$
  are identical both for Majorana or Dirac neutrinos, irrespective of
  the choice of the set of generalized CP symmetries imposed on either
  the charged leptons or neutrinos.
\item For Majorana neutrinos, the imposition of a given generalized CP
  symmetry leads to interesting correlations between the Majorana
  phases, as we will discuss later. In contrast, if neutrinos are
  Dirac fields, the Majorana phases are unphysical and can be rotated
  away by appropriate field redefinitions.

\item The imposition of generalized CP symmetries for Majorana
  neutrinos leads to important implications for neutrinoless double
  beta decays, as we discuss in Section~\ref{sec:mee}~\footnote{
    Majorana phase correlations from generalized $\mu - \tau$
    symmetry acting on neutrinos, along with their implications for
    neutrinoless double beta decay, have been investigated in
    Ref.~\cite{Chen:2015siy}.}.  In contrast, if neutrinos are Dirac
  in nature then neutrinoless double beta decay is simply forbidden.
 \end{itemize}

 In what follows we will primarily discuss the implications of
 generalized CP symmetries acting on charged leptons, assuming
 neutrinos to be Majorana particles.
 All of the resulting predictions for neutrino oscillation parameters
 hold equally well if neutrinos are Dirac--type.  In other words, all
 of the correlations between oscillation observables remain unchanged
 irrespective of whether neutrinos are Majorana or Dirac--type.
 In the latter case the Majorana phases are unphysical and
 neutrinoless double beta decay is forbidden.

%%%%%%%%%%%%%%%%%%%%%%%%%%%%%%%%%%%%%%%%%%%%%%%%%%%%%%%%%%%%%%%%%%%%%%

\section{\label{sec:main_cont}Imposing a particular CP symmetry}

%%%%%%%%%%%%%%%%%%%%%%%%%%%%%%%%%%%%%%%%%%%%%%%%%%%%%%%%%%%%%%%%%%%%%%

In this section we consider the implications of the mass matrix having
symmetry under certain generalized CP transformations, in the same
spirit as the well studied case of generalized $\mu-\tau$
symmetry~\cite{Chen:2015siy} for the neutrino sector.  From now on we
will focus in the basis in which the neutrino mass matrix is diagonal
and the CP symmetry is imposed in the charged lepton sector.

% \sout{In summary, we will assume that:}
% %
% \begin{itemize}
%  \item \sout{Neutrinos are Majorana in nature.}
%  \item \sout{We work in the basis where the neutrino mass matrix is diagonal.}
%  \item \sout{In the diagonal neutrino mass basis, we will impose a particular generalized CP symmetry on the charged leptons and study its implications.}
% \end{itemize}
%
As already explained, all of our predictions for neutrino oscillation
parameters, namely the mixing angles and $\delta_{CP}$, would remain
unchanged should neutrinos be Dirac type.
Moreover, in the Majorana case we obtain predictions for neutrinoless
double beta decay, that would be absent in the case of Dirac
neutrinos.

%%%%%%%%%%%%%%%%%%%%%%%%%%%%%%%%%%%%%%%%%%%%%%%%%%%%%%%%%%%%%%%%%%%%%%

\subsection{\label{subsec:e-mu}Generalized $e-\mu$ reflection symmetry for charged leptons}

%%%%%%%%%%%%%%%%%%%%%%%%%%%%%%%%%%%%%%%%%%%%%%%%%%%%%%%%%%%%%%%%%%%%%%

In analogy with the generalized $\mu-\tau$ symmetry in the neutrino
sector one can consider other possibilities, for example, the case of
a $e-\mu$ symmetry imposed on the charged leptons.
The CP transformation corresponding to the generalized $e-\mu$
symmetry is given by
\begin{eqnarray}
X_{e\mu}=
\left(\begin{array}{ccc}
 e^{i \alpha } \cos \Theta & i e^{\frac{1}{2} i (\alpha +\beta )} \sin \Theta & 0 \\
 i e^{\frac{1}{2} i (\alpha +\beta )} \sin \Theta & e^{i \beta } \cos \Theta & 0 \\
 0 & 0 & 1
\end{array}\right)\,,
\label{e-mu-cp-the}
\end{eqnarray}
where the three CP labels $\alpha$, $\beta$ and $\Theta$ characterize a given CP transformation. The phases $\alpha$ and $\beta$ can take any value between 0 and $2\pi$, and $\Theta$ can be limited in the range $0\leq\Theta\leq\pi$ without loss of generality. Notice that the $X_{33}$ entry can be taken to be real without loss of generality, as a global phase in the CP matrix will be unphysical. The Takagi factorization for $X_{e\mu}$ gives us
\begin{eqnarray}
\Sigma_{e\mu}=
\left(\begin{array}{ccc}
e^{i\frac{\alpha}{2}} & 0 & 0 \\
0 & e^{i\frac{\beta}{2}} & 0 \\
0 & 0 & 1
\end{array}\right) \cdot
\left(\begin{array}{ccc}
 \cos\frac{\Theta}{2} & i\sin\frac{\Theta}{2} & 0 \\
i\sin\frac{\Theta}{2} &  \cos\frac{\Theta}{2} & 0 \\
0 & 0 & 1
\end{array}\right)\,
\end{eqnarray}
Given the above assumptions and using Eq.~\eqref{eq:ugen}, the charged
lepton mixing matrix takes the form
\begin{equation}
\label{clE - Mu}U_{l}=\Sigma_{e\mu} %\sout{O_3^\dagger}
O_3^T P^{-\frac{1}{2}}\,.
\end{equation}
Since we are in the diagonal neutrino mass basis, the lepton mixing
matrix $U$ arises solely from the charged lepton sector, and it is
simply given by
\begin{eqnarray}
U\,=\,U^\dagger_l=P^\frac{1}{2} O_3 \Sigma_{e \mu}^\dagger\,,
\label{eq:PMNS_emu}
\end{eqnarray}
where the phases in the matrix $P$ can be absorbed into the charged
lepton fields. It follows that, for a given CP transformation
$X_{e\mu}$, the lepton mixing matrix i.e. all mixing angles and
CP phases depend on three free parameters $\theta_{1,2,3}$ plus the
three CP labels.
We first discuss the interesting subclass of generalized $e -\mu$
symmetry in which $\alpha=\beta=0$, leaving only one label, $\Theta$.
Using the explicit form of the lepton mixing matrix in
Eq.~\eqref{eq:PMNS_emu}, the mixing and phase parameters characterizing the lepton mixing matrix can be extracted as follows
\begin{equation}
\label{eq:angles_e-mu} \sin^2\theta_{13} = \sin^2\theta_2,\qquad \sin^2\theta_{12} = \frac{1}{2} \left(1-\cos 2\theta_3 \cos \Theta\right),\qquad \sin^2\theta_{23} = \sin^2\theta_1\,,
\end{equation}
and for the CP invariants we get
\begin{eqnarray}
& & J_{CP} = -\frac{1}{4} \sin 2 \theta_1 \sin \theta_2 \cos^2\theta_2 \sin \Theta\,, \label{e-mu-jcp}
\\
& &  I_{1} = -\frac{1}{4} \cos^4\theta_2 \sin 4\theta_3 \sin \Theta \,, \label{e-mu-i1}
\\
& & I_{2} = \frac{1}{8} \sin^22\theta_2 \sin 2\theta_3 \sin \Theta\,, \label{e-mu-i2}
\end{eqnarray}
which implies the following CP violation phases
\begin{eqnarray}
& & \tan\delta_{CP} = -\csc 2\theta_3 \tan \Theta\,,\label{e-mu-delcp}\\
& & \sin 2 \phi_{12} = -\frac{\sin 4\theta_3 \sin \Theta}{\cos^22\theta_3 \cos^2\Theta-1}\,,\label{e-mu-ph12}
\\
& & \sin 2 \phi_{13} = -\frac{\sin 2\theta_3 \sin \Theta}{\cos 2\theta_3 \cos \Theta+1}\,.\label{e-mu-ph13}
\end{eqnarray}
For the most general case, in which $\alpha\neq\beta\neq0$, the
Majorana CP phases $\phi_{12}$ and $\phi_{13}$ would change into
$\phi_{12}+(\alpha-\beta)/2$ and $\phi_{13}+\alpha/2$,
respectively, while all the mixing angles as well as the Dirac CP
phase $\delta_{CP}$ would remain intact.

In the simplifying case of $\alpha=\beta=0$, different models are
characterized by a single label, namely $\Theta$. Different
values of $\Theta$ correspond to different models of this class. In
the remaining part of the discussion we shall treat the generalized CP
label $\Theta$ as a free parameter of the theory, however it
should be kept in mind that it is really a label and any given
model will have a given fixed $\Theta$ value.

Contrasting the current ranges for the lepton mixing parameters
obtained from general neutrino oscillation global
fits~\cite{deSalas:2017kay} with the predicted relations in
Eq.~\eqref{eq:angles_e-mu}-Eq.~\eqref{e-mu-ph13}, one can obtain the
allowed ranges for the CP label $\Theta$ and the parameters
$\theta_1, \theta_2, \theta_3$.
This can in turn be used to obtain predictions for the values of the
Majorana CP phases $\phi_{12}, \phi_{13}$ determining the neutrinoless
double beta decay amplitude.

For example, taking the current best fit values of the lepton mixing
angles of~\cite{deSalas:2017kay}, the resulting values for the CP label $\Theta$ and the parameters $\theta_1, \theta_2, \theta_3$
and the Majorana CP phases $\phi_{12}, \phi_{13}$ are shown in
table~\ref{best-fit-e-mu_revise}, for both normal (NO) as well as
inverted (IO) mass ordering. In particular, if $\Theta$ is a rational
angle, i.e. a rational multiple of $\pi$, the experimental data on
lepton flavor mixing can be accommodated as well. For example, for
$\Theta=3\pi/8$, we have
\begin{table}[!h]
\begin{center}
\begin{tabular}{|c|c|c|c|c|c|c|c| } \hline \hline
\multicolumn{8}{|c|}{numerical benchmark for generalized $e-\mu$ reflection} \\ \hline \hline
~&~ $\Theta$ ~&~ $\theta_1$ ~&~ $\theta_2$ ~&~ $\theta_3$ ~&~ $\delta_{CP}$ ~&~ $\phi_{12}$ ~&~ $\phi_{13}$ \\\hline
\multirow{2}{*}{ NO }
& $69.0^{\circ}$ & $41.0^{\circ}$ & $8.4~\text{or}~171.6^{\circ}$ & $0.0^{\circ}$ & $270.0^{\circ}$ & $90.0^{\circ}$ & $0.0^{\circ}$ \\\cline{2-8}
& $111.0^{\circ}$ & $41.0^{\circ}$ & $8.4~\text{or}~171.6^{\circ}$ & $90.0^{\circ}$ & $270.0^{\circ}$ & $90.0^{\circ}$ & $90.0^{\circ}$ \\\hline
\multirow{2}{*}{IO }
& $69.0^{\circ}$ & $50.5^{\circ}$ & $8.4~\text{or}~171.6^{\circ}$ & $0.0^{\circ}$ & $270.0^{\circ}$ & $90.0^{\circ}$ & $0.0^{\circ}$ \\\cline{2-8}
& $111.0^{\circ}$ & $50.5^{\circ}$ & $8.4~\text{or}~171.6^{\circ}$ & $90.0^{\circ}$ & $270.0^{\circ}$ & $90.0^{\circ}$ & $90.0^{\circ}$ \\ \hline\hline
\end{tabular}
\end{center} \renewcommand{\arraystretch}{1.0}
\caption{\label{best-fit-e-mu_revise}
Numerical examples for the generalized $e-\mu$ reflection, where $\alpha=\beta=0$ is assumed. The values of $\Theta$ and $\theta_{1,2,3}$ are fixed by the best fit value of mixing angles and the favored value $\delta_{CP}=3\pi/2$~\cite{Adamson:2017gxd,Abe:2017uxa}. Notice that
the two Majorana CP phases are {\it predicted}, as shown in the table.  }
%The Majorana CP phases are determined from the $\Theta$ and $\theta_{1,2,3}$ values, as shown in the table.  }
\end{table}
\begin{eqnarray}
\nonumber&&\theta_{1}\simeq41.0^{\circ},\qquad \theta_{2}\simeq8.4^{\circ}~\text{or}~171.6^{\circ},\qquad \theta_{3}\simeq10.3^{\circ},\\
\nonumber&&\theta_{12}\simeq34.5^{\circ},\qquad  \theta_{13}\simeq8.44^{\circ},\qquad \theta_{23}\simeq41.0^{\circ}\,,\\
&&\delta_{CP}\simeq278.3^{\circ},\qquad \phi_{12}\simeq67.8^{\circ},\qquad \phi_{13}\simeq173.0^{\circ}\,,
\end{eqnarray}
and
\begin{eqnarray}
\nonumber&&\theta_{1}\simeq41.0^{\circ},\qquad \theta_{2}\simeq8.4^{\circ}~\text{or}~171.6^{\circ},\qquad \theta_{3}\simeq169.7^{\circ},\\
\nonumber&&\theta_{12}\simeq34.5^{\circ},\qquad  \theta_{13}\simeq8.44^{\circ},\qquad \theta_{23}\simeq41.0^{\circ}\,,\\
&&\delta_{CP}\simeq261.7^{\circ},\qquad \phi_{12}\simeq112.2^{\circ},\qquad \phi_{13}\simeq7.0^{\circ}\,.
\end{eqnarray}

\begin{figure}[!h]
\begin{center}
\begin{tabular}{cc} \hskip-0.6in
\includegraphics[width=0.7\linewidth]{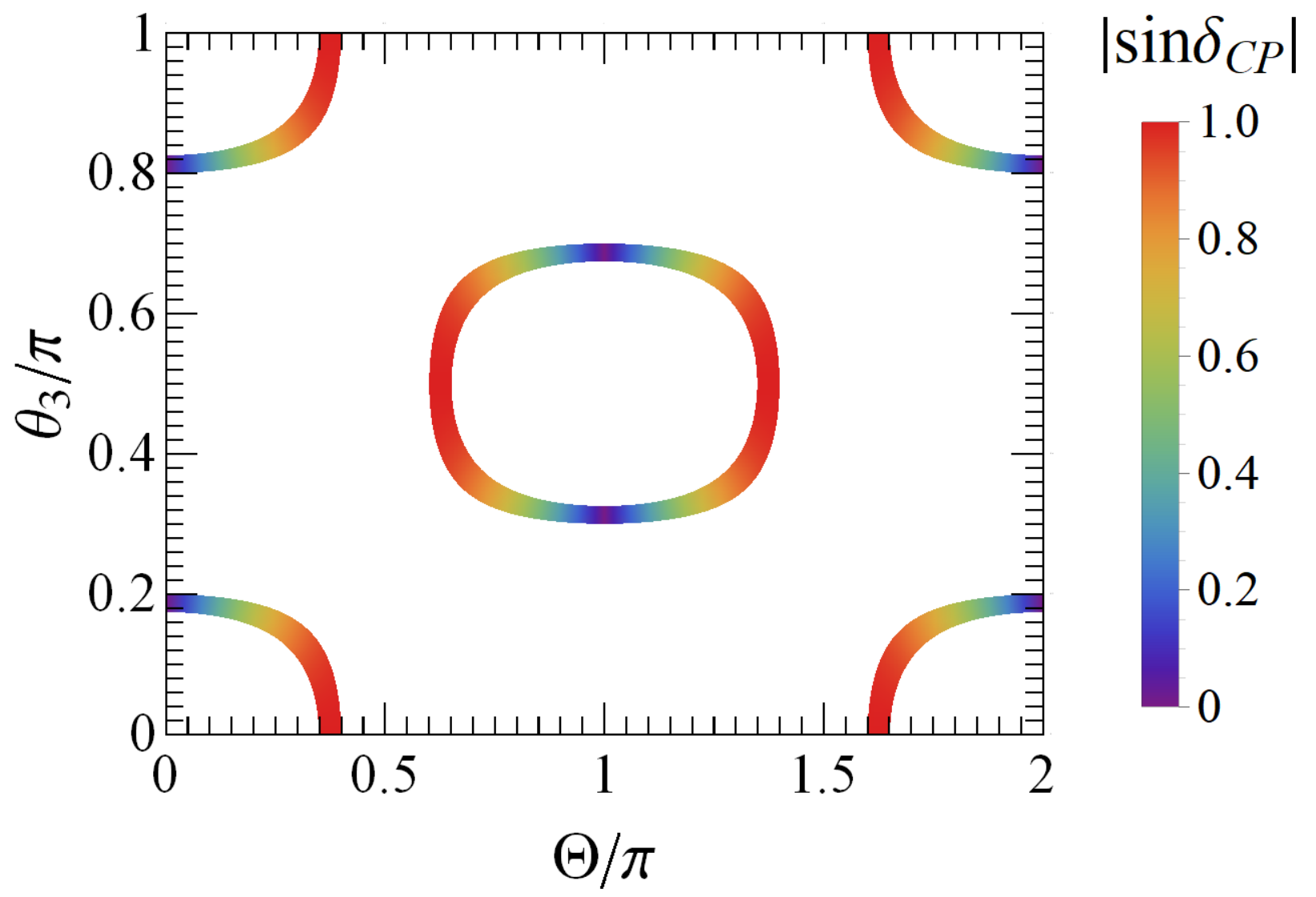}
\end{tabular}
\caption{\label{fig:theta3_Theta}Regions in the $\theta_3-\Theta$
  plane allowed by current neutrino oscillation data on
  $\sin^2\theta_{12}$ at $3\sigma$ level~\cite{deSalas:2017kay}. The
  values of $|\sin\delta_{CP}|$ are indicated by the color shadings. }
\end{center}
\end{figure}

Moreover, Eq.~\eqref{eq:angles_e-mu} implies that the solar neutrino
mixing angle $\theta_{12}$ depends on the generalized CP label
$\Theta$ and the angle $\theta_3$ of the rotation matrix $O_3$. Using
the current experimental $3\sigma$ range of the angle $\theta_{12}$
from~\cite{deSalas:2017kay}, one can obtain the allowed regions for
the $\Theta$ and $\theta_3$, as shown in
figure~\ref{fig:theta3_Theta}. To be more specific, from
Eq.~\eqref{eq:angles_e-mu} we have
\begin{eqnarray}
\cos\Theta~{\cos2\theta_3} = \cos2\theta_{12}~.
\end{eqnarray}
Thus, the value of $\Theta$ cannot be arbitrary and is restricted by
the experimental measurement of $\theta_{12}$. Inputting the 3$\sigma$
experiment range $0.273<\sin^2\theta_{12}<0.379$ given in
\cite{deSalas:2017kay}, we find that $\cos\Theta$ is constrained to be
\begin{eqnarray}
\label{eq:Theta-e-mu}
0.242<|\cos\Theta|\leq 1 \,.
\end{eqnarray}
Notice that a residual CP symmetry characterized by $\Theta=\pi/2$ is
disfavored by current oscillation data.
Since the Dirac phase $\delta_{CP}$ depends on the CP label
$\Theta$ and the parameter $\theta_{3}$ as well, as shown in
Eq.~\eqref{e-mu-delcp}, we display the contour plot of
$|\sin\delta_{CP}|$ in the $\theta_{3}-\Theta$ plane in
figure~\ref{fig:theta3_Theta}, where only the values of
$|\sin\delta_{CP}|$ in the phenomenologically viable regions are
shown.

\begin{figure}[!h]
\begin{center}
\begin{tabular}{cc}
\hskip-0.6in
\includegraphics[width=0.99\linewidth]{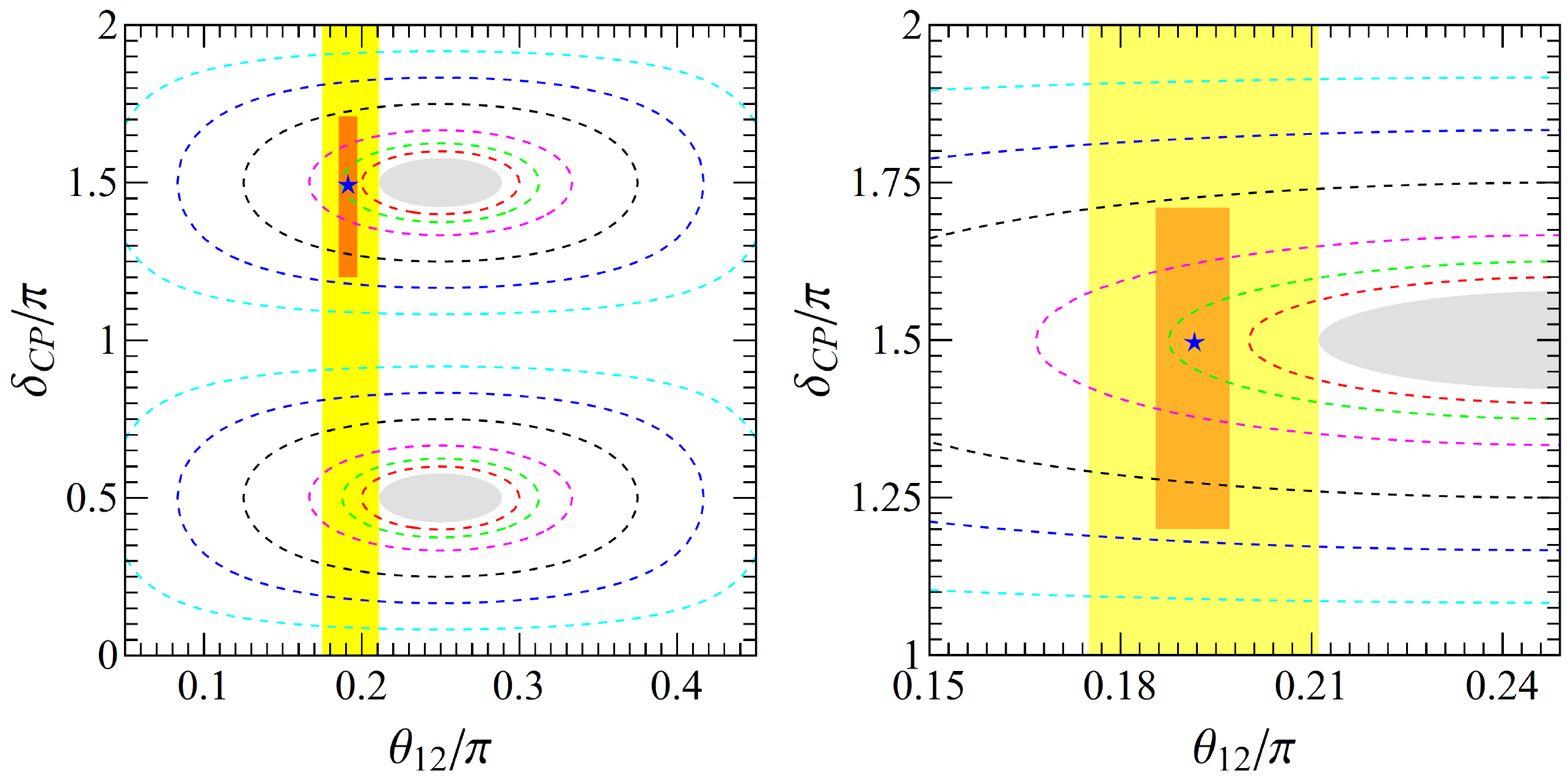}
\end{tabular}
\caption{\label{fig:s12dcp}Correlation between $\delta_{CP}$ and
  $\theta_{12}$ for various $\Theta$ values, where $\Theta$ labels the
  residual CP symmetry. Cyan, blue, black, pink, green and red dashed
  lines correspond to $\pi/12$, $\pi/6$, $\pi/4$, $\pi/3$, $3\pi/8$
  and $2\pi/5$ respectively.  The orange and yellow bands indicate the
  allowed $1\sigma$ and $3\sigma$ regions (NO case), respectively. The
  best fit point of $\theta_{12}$ and $\delta_{CP}$ is indicated with
  a star. The gray region corresponding to
  $0\leq|\cos\Theta|\leq0.242$ is excluded at the 3$\sigma$
  level~\cite{deSalas:2017kay}.}
 \end{center}
\end{figure}

Notice also that the above relations
Eq.~\eqref{eq:angles_e-mu}-Eq.~\eqref{e-mu-ph13} lead to several
important correlations among the various neutrino mixing and CP
violation parameters, e.g.
\begin{itemize}
\item{$\theta_{12}-\delta_{CP}$ correlation}
\begin{eqnarray}
\label{eq:t12_dcp}\sin^22\theta_{12}\sin^2\delta_{CP}=\sin^2\Theta\,.
\end{eqnarray}
\item{  $\phi_{12}-\delta_{CP}-\theta_{12}$ correlation}
\begin{equation}
\label{eq:sinphi12_cor_emu}\sin 2 \phi_{12}=-\frac{\cos2\theta_{12}\sin2\delta_{CP}}{1-\sin^22\theta_{12}\sin^2\delta_{CP}},\qquad \cos 2\phi_{12}=\frac{2\cos^2\delta_{CP}}{1-\sin^22\theta_{12}\sin^2\delta_{CP}}-1 \,.
\end{equation}
\item{ $ \phi_{13}-\delta_{CP}-\theta_{12}$ correlation}
\begin{equation}
\label{eq:sinphi13_cor_emu}\frac{\sin 2\delta_{CP}\sin^2\theta_{12}}{\sin2\phi_{13}}=\cos\Theta,\quad \text{or} \quad \sin^2 2\phi_{13}=\frac{\sin^22\delta_{CP}\sin^4\theta_{12}}{1-\sin^22\theta_{12}\sin^2\delta_{CP}}\,.
\end{equation}
\end{itemize}
The correlation in Eq.~\eqref{eq:t12_dcp} implies that for the
generalized $e-\mu$ symmetry in the charged lepton sector, the mixing
angle $\theta_{12}$ and the CP violating phase $\delta_{CP}$ are
correlated with each other, for a given choice of the CP label
$\Theta$, as shown in figure~\ref{fig:s12dcp}.

We see that the measurement of $\delta_{CP}$ in next generation long
baseline experiments, together with high precision measurement of
$\theta_{12}$, could help us to determine the CP label
$\Theta$ characterizing the residual CP symmetry.
Notice that this prediction is analogous to the prediction for the
atmospheric mixing angle $\theta_{23}$ and the Dirac phase
$\delta_{CP}$ obtained if we impose a generalized $\mu-\tau$
reflection in the neutrino sector~\cite{Chen:2015siy}.

Given the current experimental range of $\theta_{12}$, the other two
correlations in Eq.~\eqref{eq:sinphi12_cor_emu} and
Eq.~\eqref{eq:sinphi13_cor_emu} can be used to obtain the allowed
ranges for the Dirac CP phase $\delta_{CP}$ and the two Majorana
phases $\phi_{12}$ and $\phi_{13}$. These relations together also
imply correlations between the two Majorana phases $\phi_{12}$ and
$\phi_{13}$.

\begin{figure}[!h]
\begin{center}
\begin{tabular}{cc}
\hskip-0.6in
\includegraphics[width=0.45\linewidth]{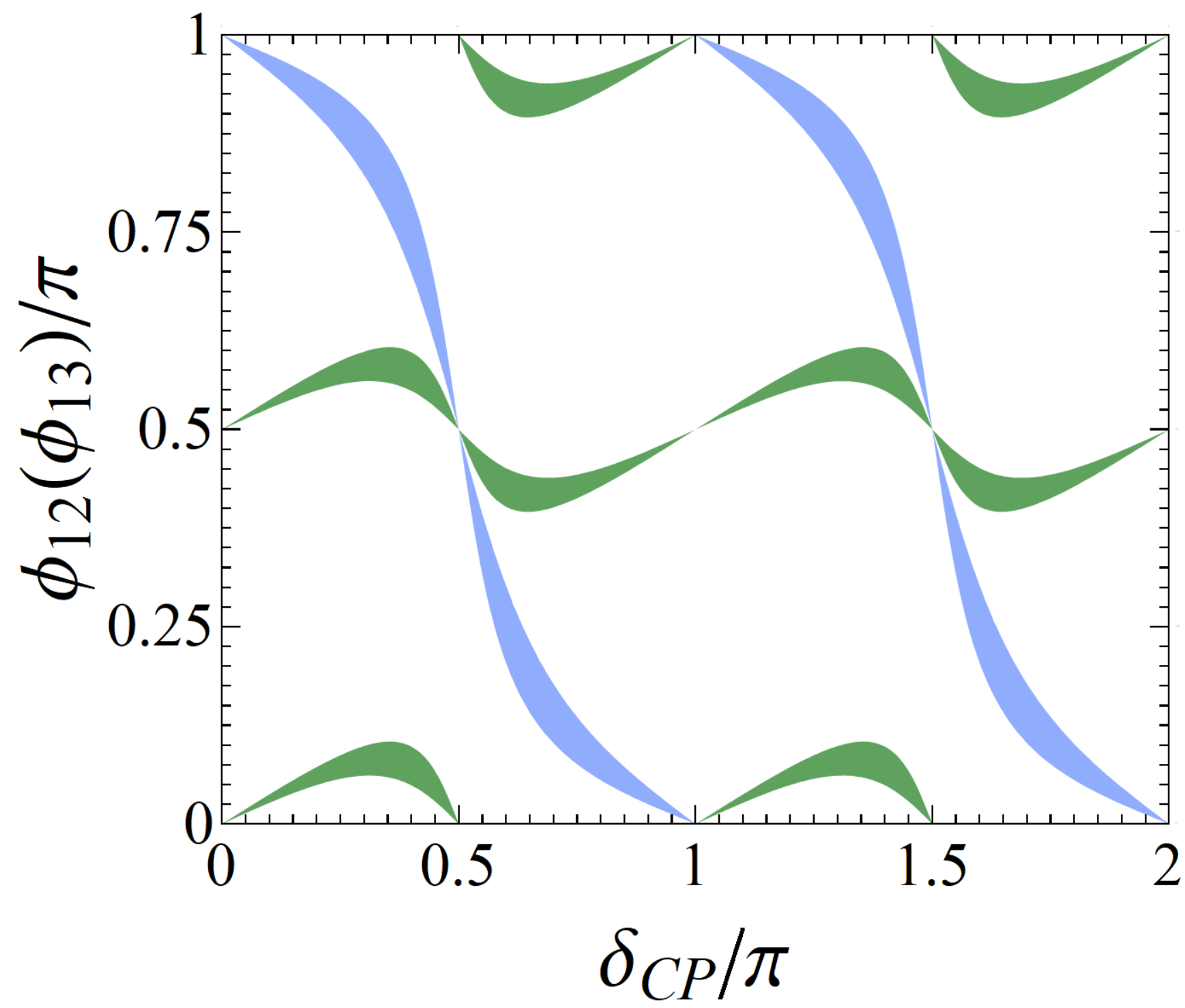}  ~~&~~
\end{tabular}
\caption{\label{fig:dcp_phi12_phi13}Correlations among the CP
  violation phases $\delta_{CP}$, $\phi_{12}$ and $\phi_{13}$ for the
  generalized $e-\mu$ reflection in the charged lepton sector, taking
  $\alpha=\beta=0$. The light blue and green areas correspond to
  $\phi_{12}-\delta_{CP}$ and $\phi_{13}-\delta_{CP}$ respectively. }
\end{center}
\end{figure}

Notice that all three parameters $\theta_{1,2,3}$ and the CP label $\Theta$ are varied randomly between 0 and $\pi$, keeping only the points for which the lepton mixing angles are consistent with
experimental data at $3\sigma$ level.
The resulting predictions for the three CP phases $\delta_{CP}$,
$\phi_{12}$ and $\phi_{13}$ are displayed in
figure~\ref{fig:dcp_phi12_phi13}. We notice that the values of
$\phi_{13}$ is around $0$, $\pi/2$ and $\pi$. The implications of the
correlations between the Majorana phases and other mixing parameters
for neutrinoless double beta decay will be further discussed in
section~\ref{sec:mee}.

%%%%%%%%%%%%%%%%%%%%%%%%%%%%%%%%%%%%%%%%%%%%%%%%%%%%%%%%%%%%%%%%%%%%%%%%%%%%%%

\subsection{\label{subsec:e-tau}Generalized $e-\tau$ reflection symmetry for charged leptons}

%%%%%%%%%%%%%%%%%%%%%%%%%%%%%%%%%%%%%%%%%%%%%%%%%%%%%%%%%%%%%%%%%%%%%%%%%%%%%%

In this section we look at the implications of imposing a generalized
$e-\tau$ symmetry on charged leptons. The corresponding CP
transformation matrix $X_{e\tau}$ is of the form~\footnote{ Note that,
although the CP label $\Theta$ characterizing the $e-\tau$
symmetry differs from that of the $e-\mu$ case, we denote it by the
same symbol (similarly also for the $\mu-\tau$ case, discussed
in subsection~\ref{subsec:mu-tau}). However, they have different
ranges of variation, see Eqs.~(\ref{eq:Theta-e-mu}),~(\ref{eq:Theta-e-tau})
and~(\ref{eq:Theta-mu-tau}).}
\begin{eqnarray}
X_{e\tau}=\left(\begin{array}{ccc}
 e^{i \alpha } \cos \Theta & 0 & i e^{\frac{1}{2} i (\alpha +\gamma )} \sin \Theta \\
 0 & 1 & 0 \\
 i e^{\frac{1}{2} i (\alpha +\gamma )} \sin \Theta & 0 & e^{i \gamma } \cos \Theta \\
\end{array}\right)\,,
\label{e-tau-cp-the}
\end{eqnarray}
where the two phase labels $\alpha$ and $\gamma$ take values
in the range $0\leq\alpha,\gamma\leq2\pi$. The angle $\Theta$ labeling
the residual CP tranformation lies in $\left[0, \pi\right]$.  The
$X_{22}$ element is taken as real without loss of generality, due to
the freedom to fix a global phase without changing the physics. We
find that the Takagi factorization of $X_{e\tau}$ is given by
\begin{eqnarray}
\Sigma_{e\tau}=
\left(\begin{array}{ccc}
e^{i\frac{\alpha}{2}} & 0 & 0 \\
0 & 1 & 0 \\
0 & 0 & e^{i\frac{\gamma}{2}}
\end{array}\right)
\left(\begin{array}{ccc}
 \cos\frac{\Theta}{2} &0 & i\sin\frac{\Theta}{2}  \\
0 &1& 0  \\
i\sin\frac{\Theta}{2} &0&  \cos\frac{\Theta}{2}
\end{array}\right)\,.
\label{sige-tau}
\end{eqnarray}
As before, we use our master formula, Eq.~\eqref{ugenn}, to extract
the total lepton mixing matrix $U$, which is the hermitian conjugate
of the charged lepton mixing matrix in the diagonal neutrino basis,
and takes the following form
\begin{eqnarray}
\label{clE - tau}U=P^\frac{1}{2} O_3 \Sigma_{e \tau}^\dagger\,.
\end{eqnarray}
In this case the mixing angles are found to be given as
\begin{eqnarray}
& &\sin^2\theta_{13} = \cos^2\theta_2 \cos^2\theta_3 \sin^2\frac{\Theta }{2}+\sin^2\theta_2 \cos^2\frac{\Theta }{2}\,,\label{e-tau-the13}
\\
& &\sin^2\theta_{12} = \frac{\sin^2\theta_3 \cos^2\theta_2}{1-\cos^2\theta_2 \cos^2\theta_3 \sin^2\frac{\Theta }{2}-\sin^2\theta_2 \cos^2\frac{\Theta }{2}}\,,\label{e-tau-the12}
\\
& &\sin^2\theta_{23} = \frac{ \sin^2\frac{\Theta }{2} \left(\sin\theta_1 \sin\theta_2 \cos\theta_3+\sin\theta_3 \cos\theta_1\right)^2+\sin^2\theta_1 \cos^2\theta_2 \cos^2\frac{\Theta }{2}}{1-\cos^2\theta_2 \cos^2\theta_3 \sin^2\frac{\Theta }{2}-\sin^2\theta_2 \cos^2\frac{\Theta }{2}}\,. \label{e-tau-the23}
\end{eqnarray}
In the limit $\alpha=\gamma=0$, the leptonic CP invariants are given by
\begin{eqnarray}
J_{CP} &=& \frac{1}{4} \sin \theta_3 \cos \theta_2 \sin \Theta \left(\sin 2 \theta_1 \left(\cos^2\theta_3-\sin^2\theta_2\sin^2\theta_3\right)+\sin \theta_2 \sin 2\theta_3 \cos 2 \theta_1\right), \nonumber \\
\,\label{e-tau-jcp} \\
I_{1} &=& \sin^2\theta_3 \cos^2\theta_2  \sin \theta_2 \cos \theta_2 \cos \theta_3 \sin \Theta  \,,\label{e-tau-I1} \\
I_{2}&=& \frac{1}{4} \Big[4 \sin \theta_2 \cos^3\theta_2 \cos^3\theta_3 \sin \Theta +4 \sin^3\theta_2 \cos \theta_2 \cos \theta_3 \sin \Theta \Big]\,. \label{e-tau-I2}
\end{eqnarray}
For the most general case $\alpha\neq \gamma \neq 0$, the Majorana CP phases satisfy
\begin{eqnarray}
\label{e-tau-phi12}\sin2\phi'_{12}&=& -\frac{\sin 2\theta_2\cos \theta_3 \sin\Theta}{2\sin^2\theta_2\sin^2\frac{\Theta}{2} +2\cos^2\theta_2\cos^2\theta_3\cos^2\frac{\Theta}{2}}\,,\\
\sin 2\phi'_{13}&=& \frac{4\sin2\theta_2\cos\theta_3\sin\Theta(\cos^2\theta_2\cos^2\theta_3-\sin^2\theta_2)}{2\cos^4\theta_2\cos^4\theta_3 \sin^2\Theta+2\sin^4\theta_2\sin^2 \Theta+\sin^22\theta_2\cos^2\theta_3(1+\cos^2\Theta)}, \nonumber
\label{e-tau-phi13} \\
\end{eqnarray}
where
\begin{equation}
2\phi_{12}^\prime=2\phi_{12}+\alpha\,, \qquad
%\sout{2\phi^\prime_{31}=2\phi_{31}+\alpha-\gamma}
~~2\phi^\prime_{13}=2\phi_{13}+\alpha-\gamma
\,.
\end{equation}
The effect of non-zero $\alpha$ and $\gamma$ is to only shift
the Majorana phases $\phi_{12}$ and $\phi_{13}$ by an amount. As in
section~\ref{subsec:e-mu}, taking the current best fit values of the
leptonic mixing parameters in~\cite{deSalas:2017kay}, the values of
the CP label $\Theta$ characterizing the CP transformation and
the $\theta_i$ can be determined, so that the Majorana CP phases
$\phi_{12}, \phi_{13}$ can be predicted if both $\alpha$ and $\gamma$ are set to be zero.

\begin{figure}[!h]
\begin{center}
\includegraphics[width=0.98\linewidth]{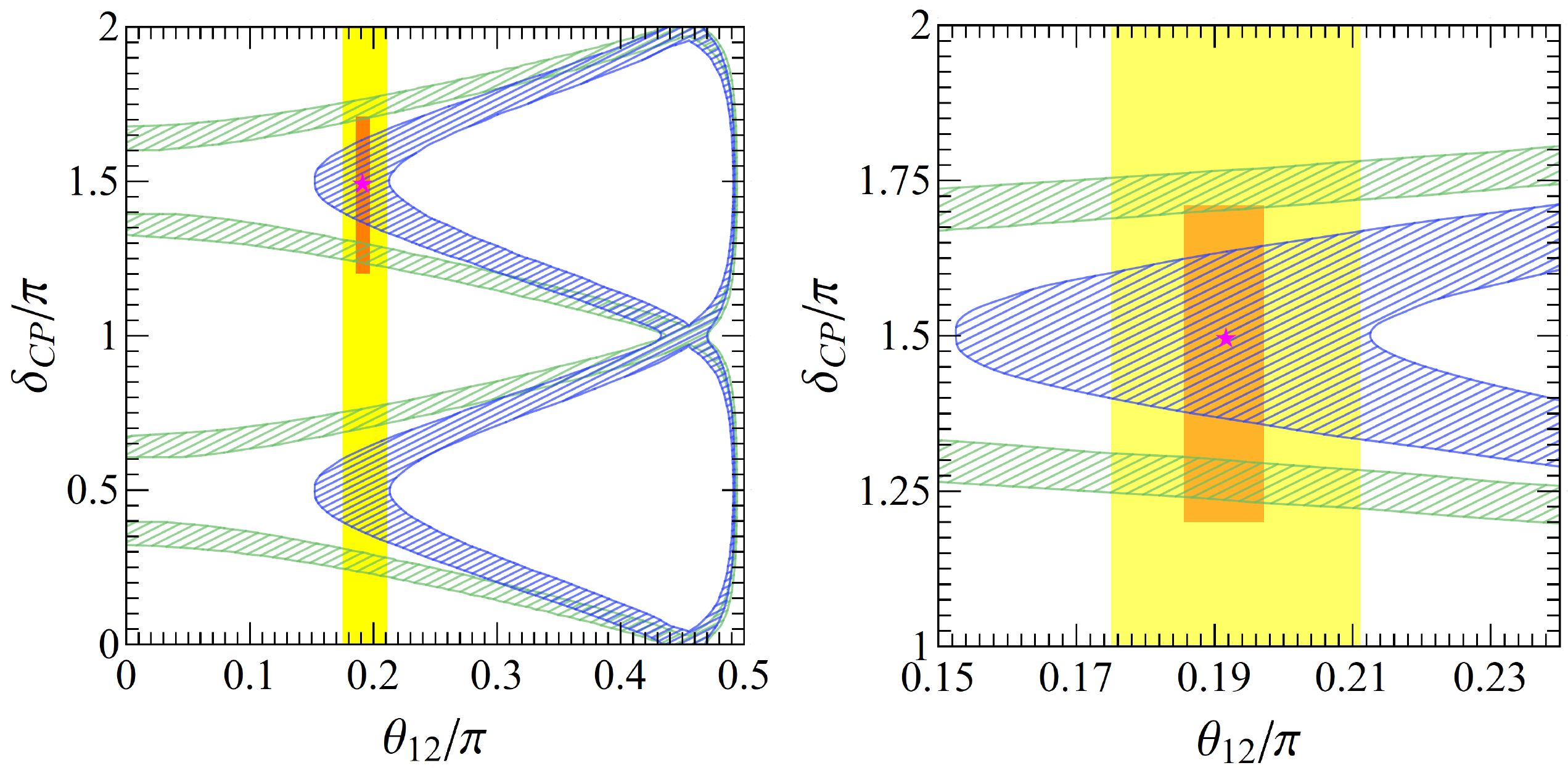}
\caption{\label{fig:etau_s12dcp} Predicted correlation between the
  Dirac CP phase $\delta_{CP}$ and the solar mixing angle
  $\theta_{12}$ in the case of generalized $e-\tau$ reflection, when
  $\theta_{13}$ and $\theta_{23}$ are in the $3\sigma$ ranges given
  in~\cite{deSalas:2017kay}. The vertical orange and yellow bands are
  the currently allowed $1\sigma$ and $3\sigma$ $\theta_{12}$ and
  $\delta_{CP}$ regions respectively for normal ordering, while the star denotes
  the best fit point. The green and light blue hatched regions
  correspond to $\Theta=\pi/12$ and $\Theta=\pi/9$ respectively. }
\end{center}
\end{figure}

The results are summarized in table~\ref{best-fit-e-tau_revise}.  As
an example, for the representative value $\Theta=\pi/9$, the best fit
values of the three lepton mixing angles~\cite{deSalas:2017kay} can be
reproduced and we have
\begin{eqnarray}
\nonumber&&\theta_{1}\simeq40.9^{\circ},\qquad \theta_{2}\simeq178.3^{\circ},\qquad \theta_{3}\simeq34.1^{\circ},\\
\nonumber&&\theta_{12}\simeq34.5^{\circ},\qquad  \theta_{13}\simeq8.44^{\circ},\qquad \theta_{23}\simeq41.0^{\circ}\,,\\
&&\delta_{CP}\simeq280.0^{\circ},\qquad \phi_{12}\simeq0.4^{\circ},\qquad \phi_{13}\simeq102.1^{\circ}\,,
\end{eqnarray}
and
\begin{eqnarray}
\nonumber&&\theta_{1}\simeq40.9^{\circ},\qquad \theta_{2}\simeq178.3^{\circ},\qquad \theta_{3}\simeq145.9^{\circ},\\
\nonumber&&\theta_{12}\simeq34.5^{\circ},\qquad  \theta_{13}\simeq8.44^{\circ},\qquad \theta_{23}\simeq41.0^{\circ}\,,\\
&&\delta_{CP}\simeq256.7^{\circ},\qquad \phi_{12}\simeq179.6^{\circ},\qquad \phi_{13}\simeq77.9^{\circ}\,.
\end{eqnarray}
\begin{table}[!h]
%\addtolength{\tabcolsep}{-2pt}
\begin{center}
\begin{tabular}{|c|c|c|c|c|c|c|c| } \hline \hline
\multicolumn{8}{|c|}{numerical benchmark  for generalized $e-\tau$ reflection} \\ \hline \hline
~&~  $\Theta$ ~&~ $\theta_1$  ~&~  $\theta_2$ ~&~ $\theta_3$ ~&~ $\delta_{CP}$  ~&~ $\phi_{12}$ ~&~   $\phi_{13}$~ \\\hline
\multirow{4}{*}{ NO }
& $20.4^{\circ}$& $40.9^{\circ}$ & $179.8^{\circ}$ & $34.1^{\circ}$ & $270.0^{\circ}$ & $0.1^{\circ}$ & $91.7^{\circ}$ \\\cline{2-8}
& $20.4^{\circ}$& $139.1^{\circ}$ & $0.2^{\circ}$ & $145.9^{\circ}$ & $270.0^{\circ}$ & $0.1^{\circ}$ & $91.7^{\circ}$ \\\cline{2-8}
& $159.6^{\circ}$& $49.5^{\circ}$ & $55.9^{\circ}$ & $89.6^{\circ}$ & $270.0^{\circ}$ & $90.1^{\circ}$ & $91.7^{\circ}$ \\\cline{2-8}
& $159.6^{\circ}$& $130.5^{\circ}$ & $124.1^{\circ}$ & $90.4^{\circ}$ & $270.0^{\circ}$ & $90.1^{\circ}$ & $91.7^{\circ}$ \\\hline
\multirow{4}{*}{IO }
& $20.3^{\circ}$& $50.7^{\circ}$ & $179.7^{\circ}$ & $145.9^{\circ}$ & $270.0^{\circ}$ & $179.9^{\circ}$ & $87.6^{\circ}$ \\\cline{2-8}
& $20.3^{\circ}$& $129.3^{\circ}$ & $0.3^{\circ}$ & $34.1^{\circ}$ & $270.0^{\circ}$ & $179.9^{\circ}$ & $87.6^{\circ}$ \\\cline{2-8}
& $159.7^{\circ}$& $38.8^{\circ}$ & $55.9^{\circ}$ & $90.6^{\circ}$ & $270.0^{\circ}$ & $89.9^{\circ}$ & $87.6^{\circ}$ \\\cline{2-8}
& $159.7^{\circ}$& $141.2^{\circ}$ & $124.1^{\circ}$ & $89.4^{\circ}$ & $270.0^{\circ}$ & $89.9^{\circ}$ & $87.6^{\circ}$ \\\hline\hline
\end{tabular}
\end{center} \renewcommand{\arraystretch}{1.0}
\caption{\label{best-fit-e-tau_revise}
Numerical examples for the generalized $e-\tau$ reflection, assuming $\alpha = \gamma = 0$. The values of $\Theta$ and $\theta_{1,2,3}$ are determined from the best fit values of neutrino mixing angles including $\delta_{CP}=3\pi/2$~\cite{Adamson:2017gxd,Abe:2017uxa}. Notice that the two Majorana CP phases are {\it predicted}, as shown in the table.  }
\end{table}
Since the lepton mixing matrix $U$ depends on three free rotation
angles $\theta_{1,2,3}$ besides the CP label $\Theta$, the mixing
angles plus CP phases are correlated with each other. After tedious
algebra calculations, we find the following relations:
\footnotesize
\begin{eqnarray}
\label{eq:etaucor1}\sin^2\Theta  & = & \frac{4s^2_{13}c^2_{13}s^2_{23}c^2_{23}c^2_{12}\sin ^2\delta _{CP}}{
s^2_{23}c^2_{23}(c^2_{12}+s^2_{12}s^2_{13})^2\sin^2\delta _{CP}+\big(s_{23}c_{23}(c^2_{12}-s^2_{12} s^2_{13})\cos\delta_{CP}+s_{13}s_{12}c_{12}\cos2\theta _{23}\big)^2}\,,\\
\sin^22 \phi_{12}  & =  & 4 s^4_{13}s^2_{23}c^2_{23}~\big( s_{23}c_{23}\left(s^2_{12}s^2_{13}-c^2_{12}\right)\cos\delta_{CP}-s_{12}c_{12}s_{13}\cos2 \theta_{23}\big)^2\sin^2\delta _{CP}\Big/
\nonumber \\
 & &  \hskip-0.1in \Big\{\big[s^2_{23}c^2_{23}(c^2_{12}+s^2_{12}s^2_{13})^2\sin^2\delta_{CP}+\big(s_{23}c_{23}(c^2_{12}-s^2_{12}s^2_{13})\cos\delta_{CP}+s_{13}s_{12}c_{12}\cos2\theta_{23}\big)^2\big]
\nonumber  \\
& &  \hskip-0.1in\times\big[s^2_{23}c^2_{23}(s^2_{13}-c^2_{12}c^2_{13})^2\sin^2\delta_{CP}+\big(s_{23}c_{23}\left(c^2_{12}-s^2_{12}s^2_{13}\right)\cos\delta _{CP}+s_{12}c_{12}s_{13}\cos2\theta _{23}\big)^2\big]\Big\}, \nonumber \\
\label{eq:etaucor4} \\
\sin2\phi_{13} &= &- \frac{2s_{23}c_{23}(c^2_{12}c^2_{13}-s^2_{13})\big(s_{23}c_{23}\left(s^2_{12}s^2_{13}-c^2_{12}\right)\cos\delta_{CP}-s_{12}c_{12}s_{13}\cos2\theta_{23}\big)\sin\delta _{CP}}{s^2_{23}c^2_{23}(s^2_{13}-c^2_{12}c^2_{13})^2\sin^2\delta_{CP}+\big(s_{23}c_{23}\left(c^2_{12}-s^2_{12}s^2_{13}\right)\cos\delta _{CP}+s_{12}c_{12}s_{13}\cos2\theta _{23}\big)^2}, \nonumber\\
\label{eq:etaucor2}
\end{eqnarray}
\normalsize
where $c_{ij}=\cos\theta_{ij}$ and $s_{ij}=\sin\theta_{ij}$. These
expressions are not very illuminating. However one can extract simple
approximations which provide a rough insight into what is going on. By
expanding the right-handed side of these equations in terms of the
small reactor angle $\theta_{13}$, one obtains
\begin{eqnarray}
\label{eq:e-tau_corr1}\sin^2\Theta &\simeq & \frac{4\sin^2\theta_{13}\sin^2\delta_{CP}}{\cos^2\theta_{12}}\left[1-4\sin\theta_{13}\tan\theta_{12} \cot2\theta_{23}\cos\delta_{CP}\right]\,, \\
\sin^22\phi_{12} &\simeq& \frac{2\sin^4\theta_{13}\sin2\delta_{CP}\sin\delta_{CP}}{\cos^4\theta_{12}} \left[\cos\delta_{CP}-4\sin\theta_{13}\tan\theta_{12}\cot2\theta_{23}\cos2\delta_{CP}\right] \nonumber\\
\label{eq:e-tau_corr2}\\
\label{eq:e-tau_corr3}\sin2\phi_{13} &\simeq & \sin 2\delta_{CP}-4\sin\theta_{13}\tan\theta_{12}\cot2\theta_{23}\sin\delta_{CP}\cos2\delta_{CP}\, .
\end{eqnarray}
One sees from Eq.~\eqref{eq:e-tau_corr1} that $\sin^2\Theta$ should be
quite small in this case as it is proportional to
$\sin^2\theta_{13}$ at leading order. Detailed numerical analysis
shows that the phenomenologically viable ranges of $\Theta$ are
\begin{equation}
\label{eq:Theta-e-tau}
\Theta\in\left[0, 0.12\pi\right]\cup\left[0.88\pi,\pi\right]\,.
\end{equation}
Note that all three CP phases would be trivial for $\Theta=0,\, \pi$.
Moreover, one sees that the solar mixing angle $\theta_{12}$ and the
Dirac CP phase $\delta_{CP}$ are strongly correlated. For example,
taking $\Theta=\pi/12$ and $\Theta=\pi/9$, the allowed regions of
$\delta_{CP}$ and $\theta_{12}$ are determined as shown in
Fig.~\ref{fig:etau_s12dcp}, where both $\theta_{13}$ and $\theta_{23}$
are required to lie within their current $3\sigma$ global
ranges~\cite{deSalas:2017kay}.

\begin{figure}[!h]
\begin{center}
\includegraphics[width=0.45\linewidth]{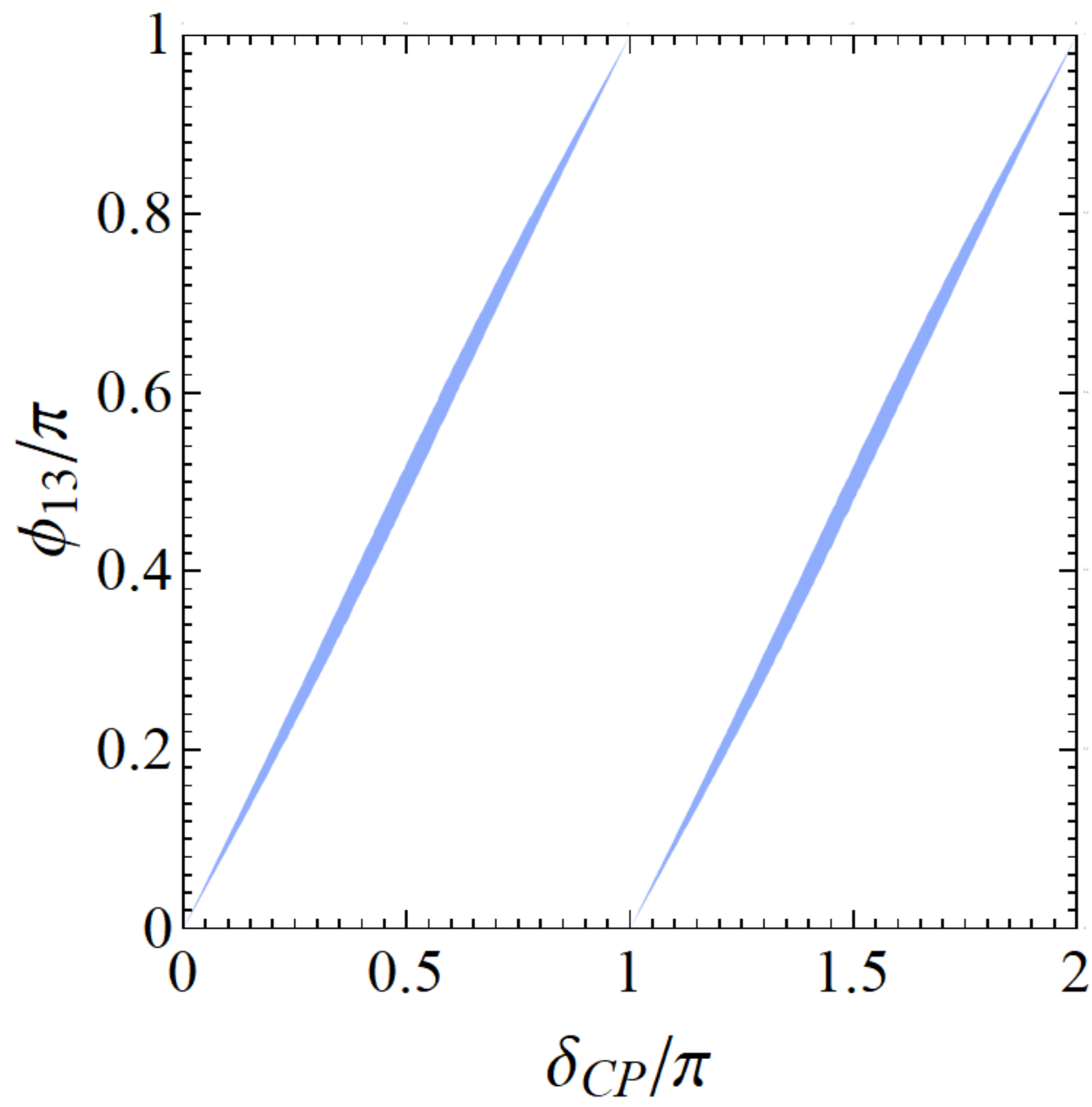}
\caption{\label{fig:phi13VSdelta_eTau} Strong correlation between
  $\phi_{13}$ and $\delta_{CP}$ for the case of generalized $e-\tau$
  reflection in the charged lepton sector, assuming
  $\alpha = \gamma = 0$ and the three mixing angles in the
  experimentally preferred $3\sigma$ regions~\cite{deSalas:2017kay}.}
\end{center}
\end{figure}

Concerning the Majorana CP phase $\phi_{12}$, from
Eq.~\eqref{eq:e-tau_corr2} one sees that in this case
$\sin2\phi_{12}$ is of order $\sin^2\theta_{13}$, so that $\phi_{12}$
lies close to $0$, $\pi/2$ or $\pi$. Moreover,
Eq.~\eqref{eq:e-tau_corr3} indicates that the Majorana phase
$\phi_{13}$ is the same as $\delta_{CP}$ up to $\pi$ at leading
order. The strong correlation between $\phi_{13}$ and $\delta_{CP}$ is
displayed in Fig.~\ref{fig:phi13VSdelta_eTau}, as expected from the
approximation of Eq.~\eqref{eq:e-tau_corr3}.
The important implications of these correlations for neutrinoless
double beta decay experiments will be discussed in
section~\ref{sec:mee}.

%%%%%%%%%%%%%%%%%%%%%%%%%%%%%%%%%%%%%%%%%%%%%%%%%%%%%%%%%%%%%%%%%%%%%%%

\subsection{\label{subsec:mu-tau}Generalized $\mu-\tau$ reflection symmetry for charged leptons}

%%%%%%%%%%%%%%%%%%%%%%%%%%%%%%%%%%%%%%%%%%%%%%%%%%%%%%%%%%%%%%%%%%%%%%%

In this section we proceed to study the implications of imposing
generalized $\mu-\tau$ symmetry on the charged lepton sector. The
analogous case of generalized $\mu-\tau$ symmetry acting on neutrinos
has been studied previously in \cite{Chen:2015siy}. The CP
transformation corresponding to the generalized $\mu-\tau$ symmetry,
$X_{\mu\tau}$ is given by
\begin{eqnarray}
\label{eq:X_mutau}
X_{\mu\tau}=\left(\begin{array}{ccc}
 1 & 0 & 0 \\
 0 & e^{i \beta } \cos \Theta & i e^{\frac{1}{2} i (\beta +\gamma )} \sin \Theta \\
 0 & i e^{\frac{1}{2} i (\beta +\gamma )} \sin \Theta & e^{i \gamma } \cos \Theta \\
\end{array}
\right)\,,
\end{eqnarray}
where the phase labels $\beta$ and $\gamma$ can take any value
between 0 and $2\pi$, and the fundamental range for the CP label $\Theta$ is $\left[0, \pi\right]$. The Takagi factorization of $X_{\mu\tau}$ is found to be
\begin{eqnarray}
\Sigma_{\mu\tau}=
\left(\begin{array}{ccc}
1 & 0 & 0 \\
0 & e^{i\frac{\beta}{2}} & 0 \\
0 & 0 & e^{i\frac{\gamma}{2}}
\end{array}\right)
\left(\begin{array}{ccc}
1 & 0  & 0   \\
0 & \cos\frac{\Theta}{2}   & i\sin\frac{\Theta}{2}  \\
0 & i\sin\frac{\Theta}{2} & \cos\frac{\Theta}{2}
\end{array}\right)\,.
\end{eqnarray}
As a result, the total lepton mixing matrix $U$ in the diagonal
neutrino basis is simply given by
\begin{eqnarray}
\label{eq:Mu-Tau}U =P^\frac{1}{2} O_3 \Sigma_{\mu\tau}^\dagger\,.
\end{eqnarray}
Notice that the matrix $\Sigma_{\mu\tau}$ is related to
$\Sigma_{e\tau}$ as follows,
\begin{equation}
\label{e-tau-mu-tau}\Sigma_{\mu\tau} = P_{12} \Sigma_{e\tau}(\alpha \rightarrow\beta) P_{12}\,,
\end{equation}
where $\alpha\rightarrow\beta$ implies replacement of phase
$\alpha$ by $\beta$ in Eq.~\eqref{sige-tau} and
$P_{12}$ is a permutation matrix given by
\begin{eqnarray}
P_{12}=
\left(\begin{array}{ccc}
0 ~&~ 1 ~&~ 0 \\
1 ~&~ 0 ~&~ 0\\
0 ~&~ 0 ~&~ 1 \end{array}\right)\,.
\end{eqnarray}
Thus it is straightforward to see that the mixing parameters predicted
by the generalized $e-\tau$ and $\mu-\tau$ symmetries obey the
following relations
\begin{eqnarray}
\label{eq:theta13_mutau}\theta_{13}^{\mu\tau}&=&\theta_{13}^{e\tau}(\theta_3\rightarrow\theta_3-\pi/2,\alpha\rightarrow\beta)\,,\\
\label{eq:theta23_mutau}\theta_{23}^{\mu\tau}&=&\theta_{23}^{e\tau}(\theta_3\rightarrow\theta_3-\pi/2,\alpha\rightarrow\beta)\,, \\
\label{eq:theta12_mutau}\theta_{12}^{\mu\tau}&=&\pi/2-\theta_{12}^{e\tau}(\theta_3\rightarrow\theta_3-\pi/2,\alpha\rightarrow\beta)\,,\\
\label{eq:deltaCP_mutau}\delta_{CP}^{\mu\tau}&=&\delta_{CP}^{e\tau}(\theta_3\rightarrow\theta_3-\pi/2,\alpha\rightarrow\beta)+\pi\,,\\
\label{eq:phi12_mutau}\phi_{12}^{\mu\tau}&=&-\phi_{12}^{e\tau}(\theta_3\rightarrow\theta_3-\pi/2,\alpha\rightarrow\beta)\,,\\
\label{eq:phi13_mutau}\phi_{13}^{\mu\tau}&=&\phi_{13}^{e\tau}(\theta_3\rightarrow\theta_3-\pi/2,\alpha\rightarrow\beta)-\phi_{12}^{e\tau}(\theta_3\rightarrow\theta_3-\pi/2,\alpha\rightarrow\beta)\,.
\end{eqnarray}
From the predictions of generalized $e-\tau$ symmetry in previous
section, the above relations can be used to obtain the values of the
mixing angles and CP phases corresponding the generalized $\mu-\tau$
symmetry.
As before, the values of the parameters $\theta_{1,2,3}$ and the CP label $\Theta$ can be determined from the measured values
of three lepton mixing angles and the favored
$\delta_{CP}\simeq3\pi/2$ value~\cite{Adamson:2017gxd,Abe:2017uxa}.

\begin{table}[!h]
\begin{center}
\begin{tabular}{|c|c|c|c|c|c|c|c| } \hline \hline
\multicolumn{8}{|c|}{numerical benchmark for generalized $\mu-\tau$ reflection} \\ \hline \hline
 ~&~ $\Theta$ ~&~ $\theta_1$ ~&~ $\theta_2$ ~&~ $\theta_3$ ~&~ $\delta_{CP}$ ~&~  $\phi_{12}$ ~&~ $\phi_{13}$  \\\hline
\multirow{4}{*}{NO }
& $29.3^{\circ}$ & $40.6^{\circ}$ & $0.5~\text{or}~179.5^{\circ}$ & $144.6^{\circ}$ & $270.0^{\circ}$ & $0.2^{\circ}$ & $86.5^{\circ}$ \\\cline{2-8}
& $29.3^{\circ}$ & $139.4^{\circ}$ & $0.5~\text{or}~179.5^{\circ}$ & $35.4^{\circ}$ & $270.0^{\circ}$ & $0.2^{\circ}$ & $86.5^{\circ}$ \\\cline{2-8}
& $150.7^{\circ}$ & $49.8^{\circ}$ & $144.6~\text{or}~35.4^{\circ}$ & $179.3^{\circ}$ & $270.0^{\circ}$ & $90.2^{\circ}$ & $176.5^{\circ}$ \\\cline{2-8}
& $150.7^{\circ}$ & $130.2^{\circ}$ & $144.6~\text{or}~35.4^{\circ}$ & $0.7^{\circ}$ & $270.0^{\circ}$ & $90.2^{\circ}$ & $176.5^{\circ}$ \\\hline
\multirow{4}{*}{IO }
& $29.2^{\circ}$ & $51.1^{\circ}$ & $179.3~\text{or}~0.7^{\circ}$ & $35.4^{\circ}$ & $270.0^{\circ}$ & $179.7^{\circ}$ & $94.9^{\circ}$ \\\cline{2-8}
& $29.2^{\circ}$ & $128.9^{\circ}$ & $179.3~\text{or}~0.7^{\circ}$ & $144.6^{\circ}$ & $270.0^{\circ}$ & $179.7^{\circ}$ & $94.9^{\circ}$ \\\cline{2-8}
& $150.8^{\circ}$ & $38.4^{\circ}$ & $35.4~\text{or}~144.6^{\circ}$ & $0.9^{\circ}$ & $270.0^{\circ}$ & $89.7^{\circ}$ & $4.9^{\circ}$ \\\cline{2-8}
& $150.8^{\circ}$ & $141.6^{\circ}$ & $35.4~\text{or}~144.6^{\circ}$ & $179.1^{\circ}$ & $270.0^{\circ}$ & $89.7^{\circ}$ & $4.9^{\circ}$ \\\hline\hline
\end{tabular}
\end{center} \renewcommand{\arraystretch}{1.0}
\caption{\label{best-fit-mu-tau_revise} Numerical examples for the
eneralized $\mu-\tau$ reflection, where we assume $\beta=\gamma=0$.
The values of $\Theta$ and $\theta_{1,2,3}$ are fixed by the best
fitted value of mixing angles and experimentally favored value
$\delta_{CP}=3\pi/2$~\cite{Adamson:2017gxd,Abe:2017uxa}.
Notice that the two Majorana CP phases are {\it predicted}, as shown in the table.  }
%The Majorana CP phases are determined from the $\Theta$ and $\theta_{1,2,3}$ values, as shown in the table.  }
\end{table}
\begin{figure}[!t]
\begin{center}
\includegraphics[width=0.98\linewidth]{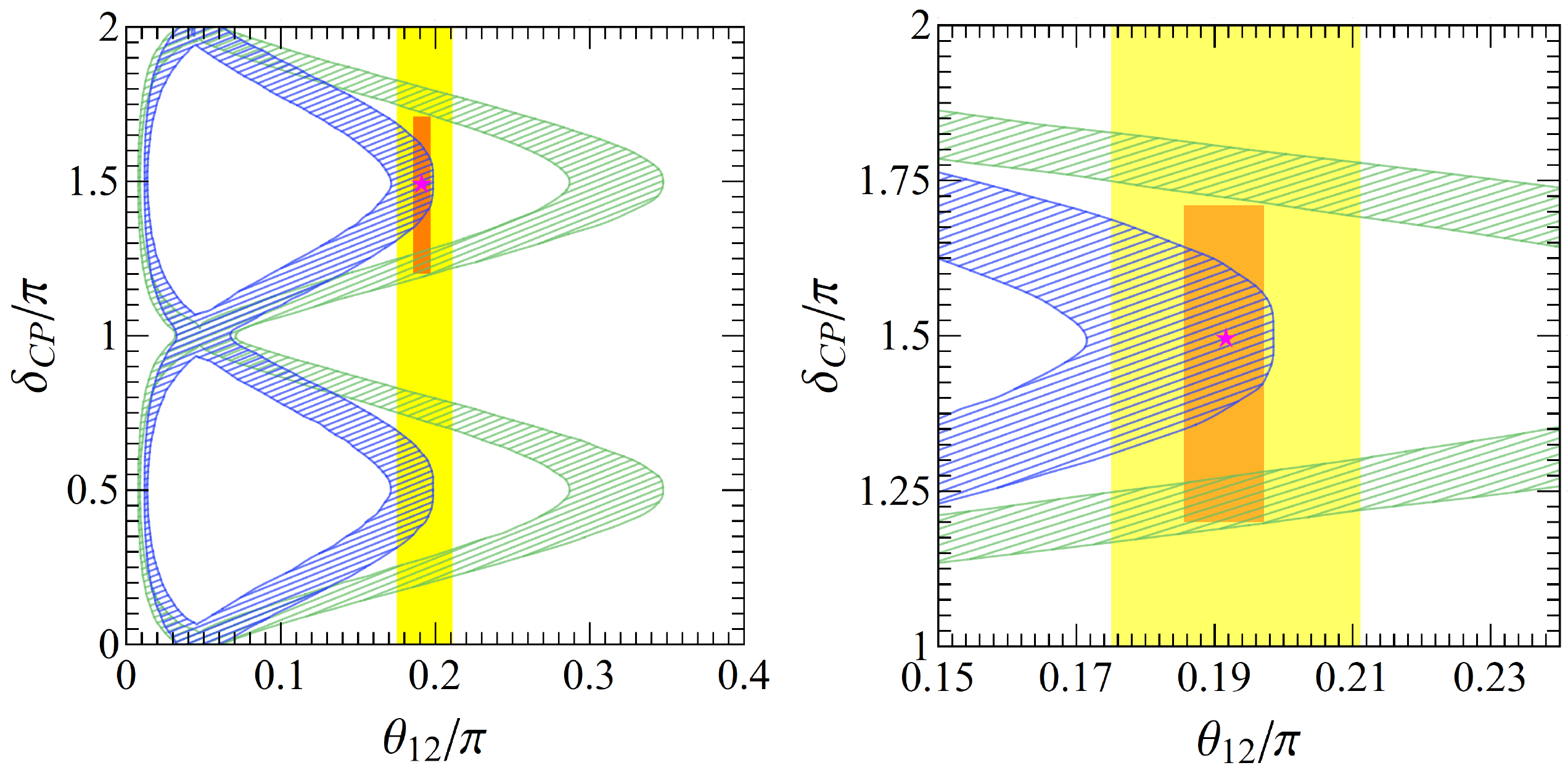}
\caption{\label{fig:mutau_s12dcp} Predicted correlation between the
  Dirac CP phase $\delta_{CP}$ as a function of the solar mixing angle
  $\theta_{12}$ in the case of generalized $\mu-\tau$ reflection, when
  $\theta_{13}$ and $\theta_{23}$ are required to lie in their
  $3\sigma$ ranges.  The vertical orange and yellow bands are the
  currently allowed $1\sigma$ and $3\sigma$ (NO case) respectively and
  the star denotes the best fit point~\cite{deSalas:2017kay}.  The
  green and light blue hatched regions correspond to $\Theta=\pi/9$
  and $\Theta=\pi/6$ respectively.}
\end{center}
\end{figure}
The predicted Majorana CP phases are given in
table~\ref{best-fit-mu-tau_revise}. As an example, taking
$\Theta=\pi/6$, we have
\begin{eqnarray}
\nonumber&&\theta_{1}\simeq40.6^{\circ} ,\qquad \theta_{2}\simeq0^{\circ},\qquad \theta_{3}\simeq145.1^{\circ},\\
\nonumber&&\theta_{12}\simeq33.9^{\circ},\qquad  \theta_{13}\simeq8.51^{\circ},\qquad \theta_{23}\simeq41.0^{\circ}\,,\\
&&\delta_{CP}\simeq273.7^{\circ} ,\qquad \phi_{12}\simeq0^{\circ},\qquad \phi_{13}\simeq90^{\circ}\,,
\end{eqnarray}
and
\begin{eqnarray}
\nonumber&&\theta_{1}\simeq40.6^{\circ} ,\qquad \theta_{2}\simeq180^{\circ},\qquad \theta_{3}\simeq34.9^{\circ},\\
\nonumber&&\theta_{12}\simeq33.9^{\circ},\qquad  \theta_{13}\simeq8.51^{\circ},\qquad \theta_{23}\simeq41.0^{\circ}\,,\\
&&\delta_{CP}\simeq273.7^{\circ} ,\qquad \phi_{12}\simeq180^{\circ},\qquad \phi_{13}\simeq90^{\circ}\,,
\end{eqnarray}

As in the previous two cases, here we also find that the relations
Eq.~\eqref{eq:theta13_mutau} - Eq.~\eqref{eq:phi13_mutau} lead to
correlations between the neutrino oscillation parameters. The exact
results are lengthy and not very illuminating. By expanding in the
small quantity $\sin\theta_{13}$, one can display the predictions in a
simple form
\begin{eqnarray}
\label{eq:corr1_mu-tau}&&\sin^2\Theta\simeq \frac{4\sin^2\theta_{13}\sin^2\delta_{CP}}{\sin^2\theta_{12}}\left[1+4\sin\theta_{13}\cot\theta_{12}\cot2\theta_{23}\cos\delta_{CP}\right]\,,\\
&&\sin^2 2\phi_{12}\simeq \frac{2\sin^4\theta_{13}\sin2\delta_{CP}\sin\delta_{CP}}{\sin^4\theta_{12}} \left[\cos\delta_{CP}+4\sin\theta_{13}\cot\theta_{12}\cot2\theta_{23}\cos2\delta_{CP}\right], \nonumber \\
\label{eq:corr2_mu-tau}\\
\label{eq:corr3_mu-tau}&&\sin 2(\phi_{13}-\phi_{12}) \simeq\sin 2\delta_{CP}+4\sin\theta_{13}\cot\theta_{12}\cot2\theta_{23}\sin\delta_{CP}\cos2\delta_{CP}\,.
\end{eqnarray}
\begin{figure}[t]
\begin{center}
\includegraphics[width=0.45\linewidth]{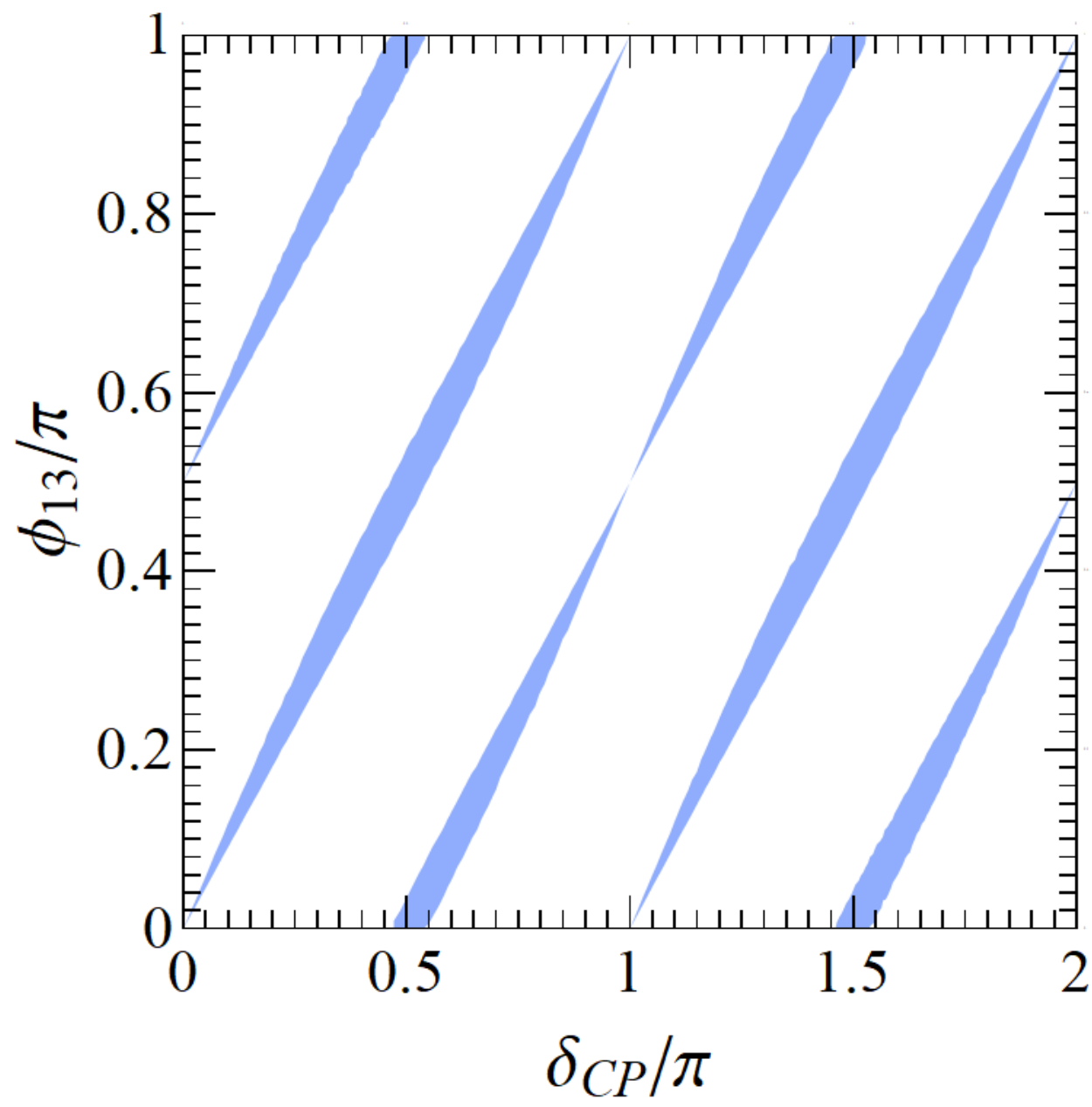}
\caption{\label{fig:phi13VSdelta_muTau} The correlation between
  $\phi_{13}$ and $\delta_{CP}$ for the generalized $\mu-\tau$
  reflection acting on the charged lepton sector, taking
  $\beta=\gamma=0$ and the three mixing angles in their
  $3\sigma$ regions~\cite{deSalas:2017kay}.}
\end{center}
\end{figure}

Taking into account the current allowed $3\sigma$ range
from~\cite{deSalas:2017kay}, Eq.~\eqref{eq:corr1_mu-tau} gives rise to
the following viable range of $\Theta$,
\begin{equation}
\label{eq:Theta-mu-tau}
\Theta\in\left[0, 0.18\pi\right]\cup\left[0.82\pi, \pi\right]\,.
\end{equation}
Notice that the relation between mixing angles and $\delta_{CP}$ in
Eq.~\eqref{eq:corr1_mu-tau} is different from those in
Eq.~\eqref{eq:t12_dcp} and Eq.~\eqref{eq:e-tau_corr1} corresponding to
the cases of generalized $e-\mu$ and $e-\tau$ reflection,
respectively.
Taking, for example, $\Theta=\pi/9$ and $\Theta=\pi/6$, the possible
values of $\delta_{CP}$ as a function of the solar mixing angle
$\theta_{12}$ are shown in Fig.~\ref{fig:mutau_s12dcp}. The
measurement of $\delta_{CP}$ in future neutrino oscillation
experiments can help us to fix the value of the CP label $\Theta$. In addition, Eq.~\eqref{eq:corr2_mu-tau} implies that
$\sin2\phi_{12}$ is quite small, so that $\phi_{12}$ lies close to
$0$, $\pi/2$ and $\pi$.
In addition, one can see from Eq.~\eqref{eq:corr3_mu-tau} that the
Majorana phase $\phi_{13}$ is close to $\delta_{CP}~(\text{mod}~\pi)$
or $\delta_{CP}+\pi/2~(\text{mod}~\pi)$, up to higher order terms in
$\sin\theta_{13}$. The predicted correlation between $\phi_{13}$ and
$\delta_{CP}$ in shown in Fig.~\ref{fig:phi13VSdelta_muTau}.
\begin{figure}[!h]
\begin{center}
\hskip-0.3in\includegraphics[width=0.5\linewidth]{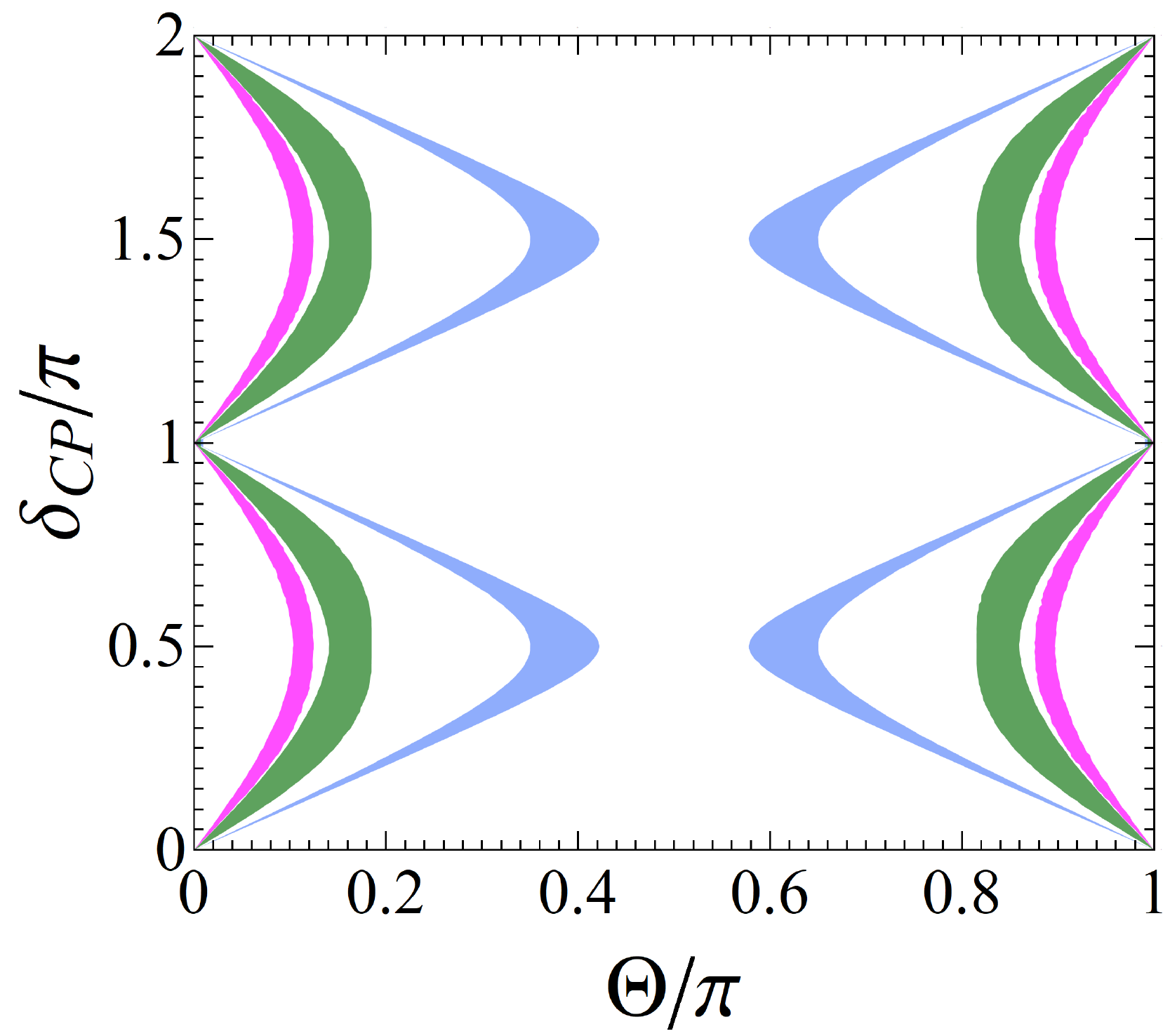}
\caption{\label{delta_cp} {Predictions for the Dirac CP violation phase $\delta_{CP}$ as a function of the CP label $\Theta$ associated with generalized CP symmetries. The light blue, magenta and green regions correspond to the cases of generalized $e-\mu$, $e-\tau$ and $\mu-\tau$ symmetries, respectively. }}
 \end{center}
\end{figure}

To conclude this section we stress that the imposition of generalized
CP symmetries implies correlations involving the lepton mixing angles,
the Dirac CP violating phase relevant for neutrino oscillations.
We have already discussed some of the phenomenological implications of
these correlations. Here we mention the generalized CP symmetry
predictions for the CP violation phase $\delta_{CP}$.

Figure \ref{delta_cp} shows how the correlations between the Dirac
phase $\delta_{CP}$ characterizing neutrino oscillations and the
CP label $\Theta$.
Note that the predictions for this phase do not depend on the values
of the phase labels $\alpha$, $\beta$ and $\gamma$. This is as expected, since oscillation probabilities can not depend on Majorana phases~\cite{Schechter:1981gk}. In principle, by fixing a given CP symmetry or $\Theta$ value one can make predictions for the upcoming generation of neutrino oscillation experiments such as DUNE~\cite{Acciarri:2016ooe,Acciarri:2015uup}. This exercise has been
carried out in the analogous case of generalized CP symmetries acting
on neutrinos in Refs.~\cite{Chen:2015siy,Chen:2016ica}.
In the remaining of this paper we prefer to focus on the rather
significant implications of the above predictions for the upcoming
neutrinoless double beta decay experiments, as discussed in the next
section.

%%%%%%%%%%%%%%%%%%%%%%%%%%%%%%%%%%%%%%%%%%%%%%%%%%%%%%%%%%%%%%%%%%%%%%%%%%%%

\section{\label{sec:mee} Implications for neutrinoless double beta decay}

%%%%%%%%%%%%%%%%%%%%%%%%%%%%%%%%%%%%%%%%%%%%%%%%%%%%%%%%%%%%%%%%%%%%%%%%%%%%

We shall proceed to investigate the implications of generalized CP
symmetries of charged leptons in theories with Majorana neutrinos,
assuming the basis in which their mass matrix is
diagonal~\footnote{For Dirac neutrinos the Majorana phases are
  unphysical, as they can be eliminated by field
  redefinitions~\cite{Schechter:1980gr,Schechter:1981gk}. Likewise, in
  this case neutrinoless double beta decay is always forbidden.}.
As in previous sections, we shall set $\alpha=\beta=\gamma=0$ in this section.

As we already saw at length, the imposition of generalized $e-\mu$,
$e-\tau$ and $\mu-\tau$ symmetries leads to correlations amongst the
lepton mixing angles, the Dirac CP violating phase $\delta_{CP}$, and
also the two Majorana phases $\phi_{12}$, $\phi_{13}$.  We have
already discussed some features of these correlations.  Here we focus
on the generalized CP symmetry predictions involving the Majorana CP
violation phases.

\begin{figure}[t]
\begin{center}
\hskip-0.3in\includegraphics[width=0.98 \linewidth]{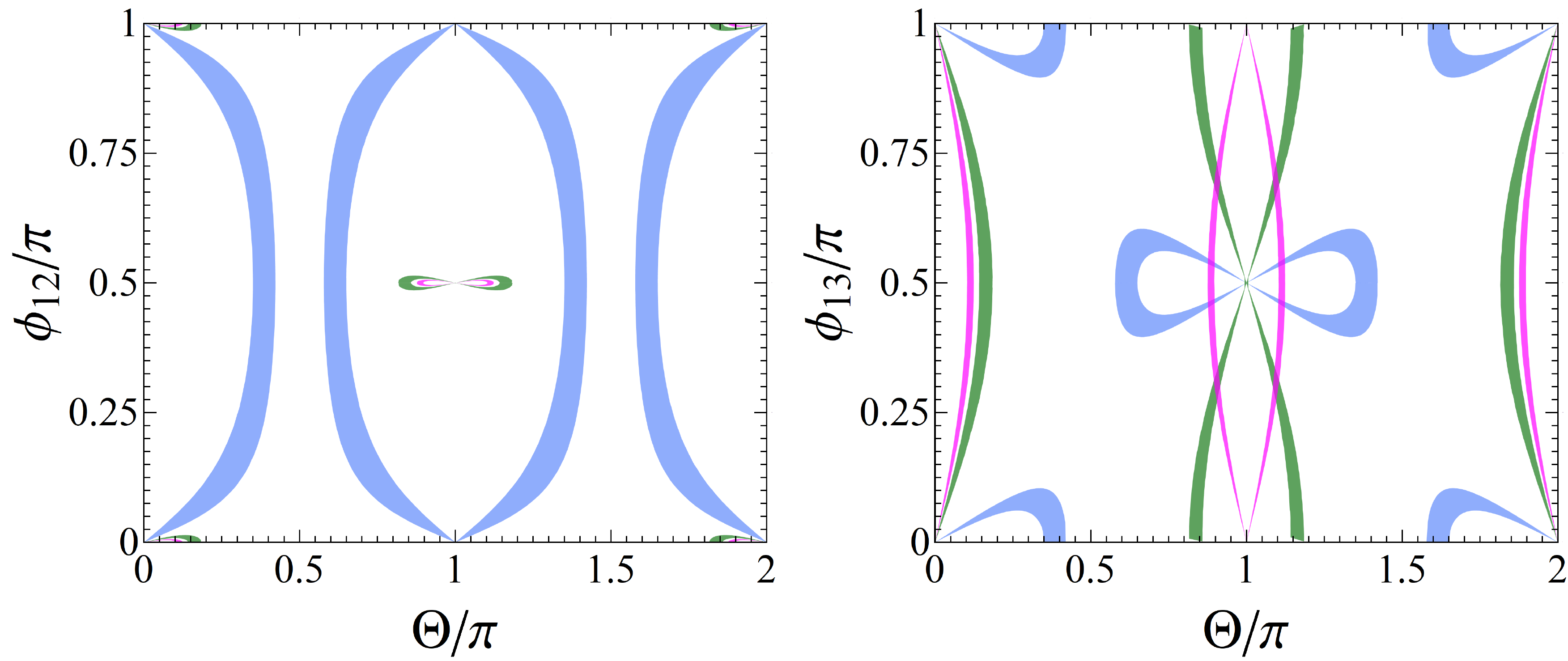}
\caption{\label{mee-the-phi}{Predictions for the Majorana CP violation phases $\phi_{12}$ and $\phi_{13}$ as a function of the
CP label $\Theta$. The light blue, magenta and green
regions are for the generalized $e-\mu$, $e-\tau$ and $\mu-\tau$
symmetries, respectively. Here we choose $\alpha = \beta=\gamma=0$.}}
 \end{center}
\end{figure}
\begin{figure}[!h]
\begin{center}
\hskip-0.3in\includegraphics[width=0.5\linewidth]{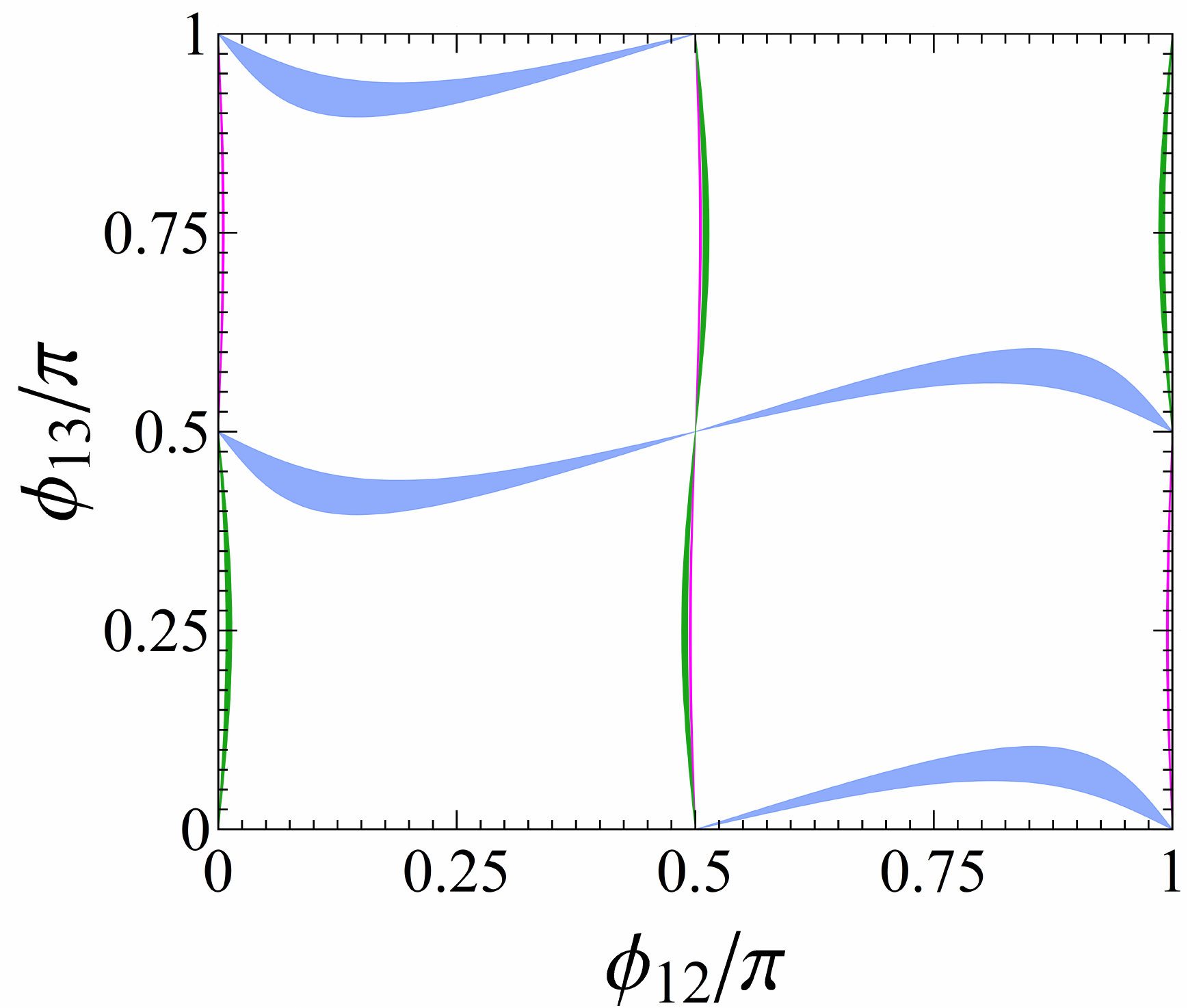}
\caption{\label{mee-phi-phi}{Correlations between the two Majorana
phases $\phi_{12}$ and $\phi_{13}$ of the lepton mixing matrix. The light blue, magenta and green regions are for the generalized $e-\mu$, $e-\tau$ and $\mu-\tau$ symmetries respectively. The magenta regions are quite narrow and hardly visible, and they are closely distributed around $\phi_{12}\sim0$, $\phi_{12}\sim\pi/2$ and $\phi_{12}\sim\pi$. Here we take $\alpha = \beta=\gamma=0$} for illustration.}
\end{center}
\end{figure}

By treating $\theta_{1,2,3}$ and the CP label $\Theta$ as
random numbers in the range of $0$ and $\pi$ we perform a
comprehensive numerical scan of the parameter space.
The three lepton mixing angles are required to lie within their
$3\sigma$ ranges, as determimed in~\cite{deSalas:2017kay}. In
Fig.~\ref{mee-the-phi} we show the correlations between the Majorana
CP phases $\phi_{12}, \phi_{13}$ with the CP label $\Theta$
characterizing the three generalized CP symmetries discussed in
previous section.
As clear from Fig.~\ref{mee-the-phi}, the Majorana phases $\phi_{12}$
and $\phi_{13}$ are strongly correlated with the generalized CP
label $\Theta$.
This in turn implies that they are correlated with other mixing
parameters as discussed in previous section, as well as mutually
correlated, as shown in Fig.~\ref{mee-phi-phi}.

\begin{figure}[!h]
\begin{center}
\begin{tabular}{c} \hskip-0.6in
\includegraphics[width=0.495\linewidth]{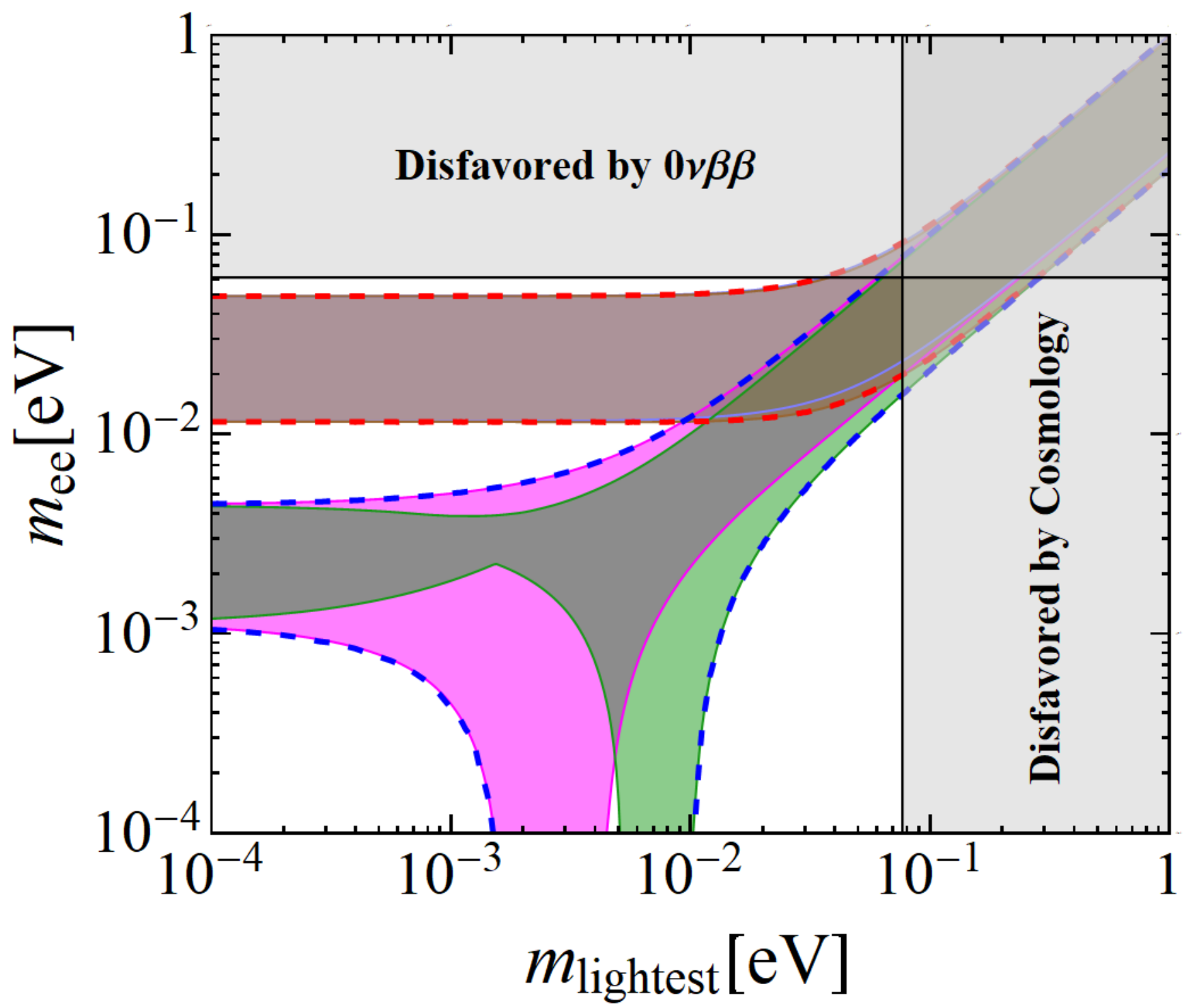}  \includegraphics[width=0.495\linewidth]{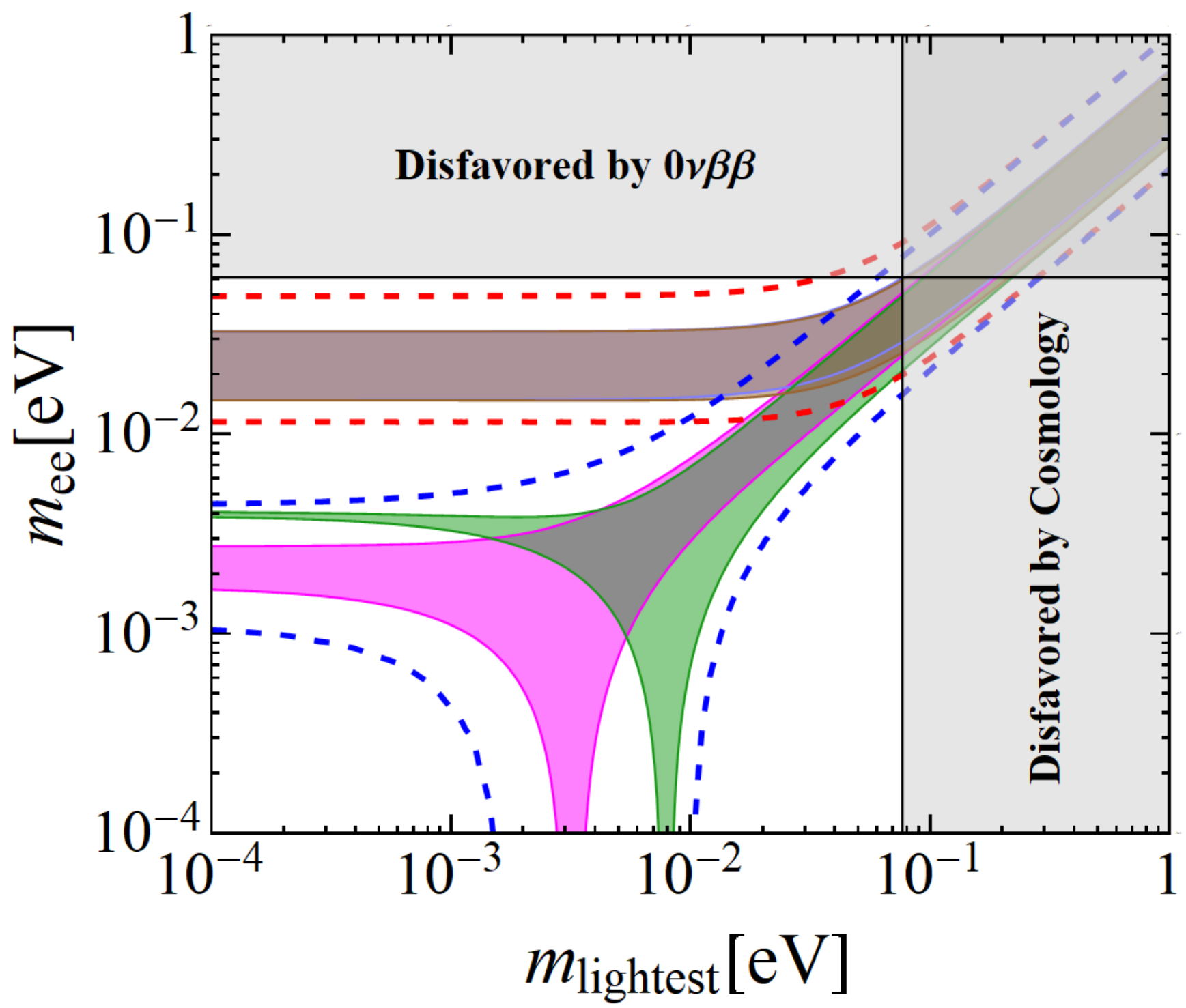}\\[0.05in]
\hskip-0.1in\includegraphics[width=0.90\linewidth]{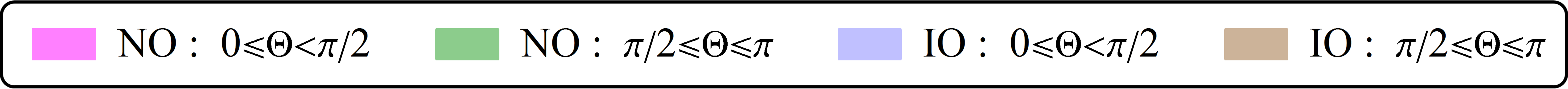}
\end{tabular}
\caption{\label{mee-e-mu} Attainable values of the effective Majorana
mass $m_{ee}$ as a function of the lightest neutrino mass in the generalized $e-\mu$ reflection scheme. In the left panel the CP label $\Theta$ is varied over all its values, while we fix
$\Theta=2\pi/5$ and $\Theta=3\pi/5$ in the right panel. Notice that
the expected $m_{ee}$ for $0\leq\Theta<\pi/2$ and $\pi/2\leq\Theta\leq\pi$ almost coincide in the case of IO. The red (blue) dashed lines delimit the most general allowed regions for IO (NO) neutrino mass spectrum obtained by varying the mixing parameters over their $3\sigma$ ranges~\cite{deSalas:2017kay}. The present most stringent upper limits $m_{ee}<0.061$ eV from KamLAND-ZEN~\cite{KamLAND-Zen:2016pfg} and EXO-200~\cite{Albert:2017owj} is shown by horizontal grey band. The vertical grey band indicates the current sensitivity of cosmological data from the Planck collaboration~\cite{Ade:2015xua}. }
\end{center}
\end{figure}

These correlations have interesting consequences for the effective
Majorana mass $m_{ee}$ characterizing the neutrinoless double beta
decay amplitude, as we discuss next.
We first consider the case of generalized $e-\mu$ reflection. The
relevant results and correlations for this case have already been
discussed in section~\ref{subsec:e-mu}. The effective Majorana mass
$m_{ee}$ characterizing the neutrinoless double beta decay amplitude
for case of generalized $e-\mu$ symmetry is shown in
Fig.~\ref{mee-e-mu}.

As is clear from this figure, almost all the $3\sigma$ range values of
the effective mass $m_{ee}$ can be reproduced if the variation of the
CP label $\Theta$ is taken into account. However, the allowed
range of $m_{ee}$ for a given fixed $\Theta$ is quite restricted, for
both NO and IO. We note that due to the constraints on mixing angles
and CP phases, the range of $m_{\text{lightest}}$ in which $m_{ee}$
can be very small for NO is considerably reduced. Along with the
predictions for the neutrino oscillation parameters, the predicted
range for $m_{ee}$ can also be used to test the generalized $e-\mu$
reflection.  We also indicate by the vertical grey band the
sensitivity $\sum m_i<0.230$ eV at $95\%$ C.L. limit claimed by the
Planck collaboration~\cite{Ade:2015xua}.

\begin{figure}[hptb!]
\begin{center}
\begin{tabular}{c} \hskip-0.6in
\includegraphics[width=0.495\linewidth]{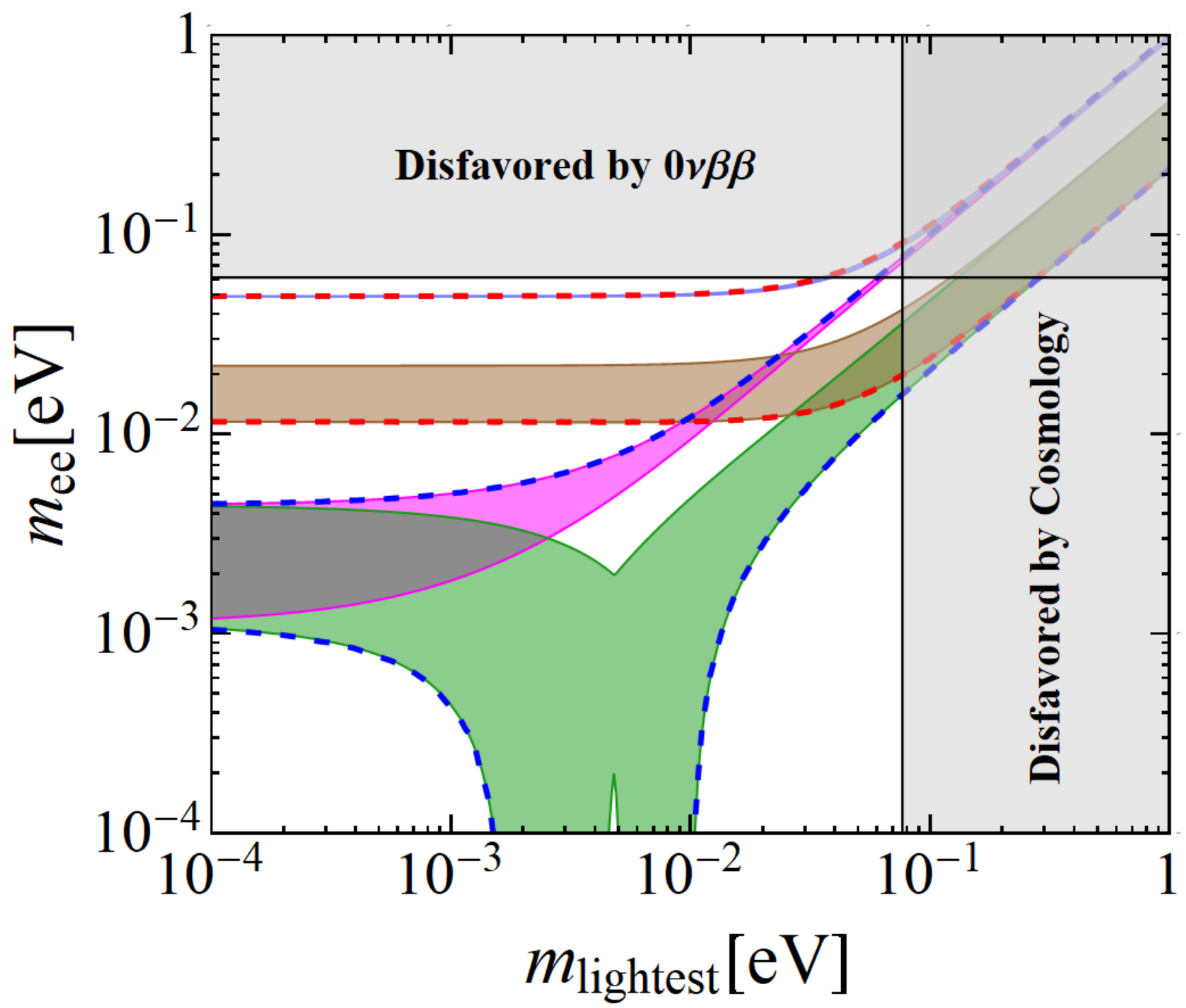}
\includegraphics[width=0.495\linewidth]{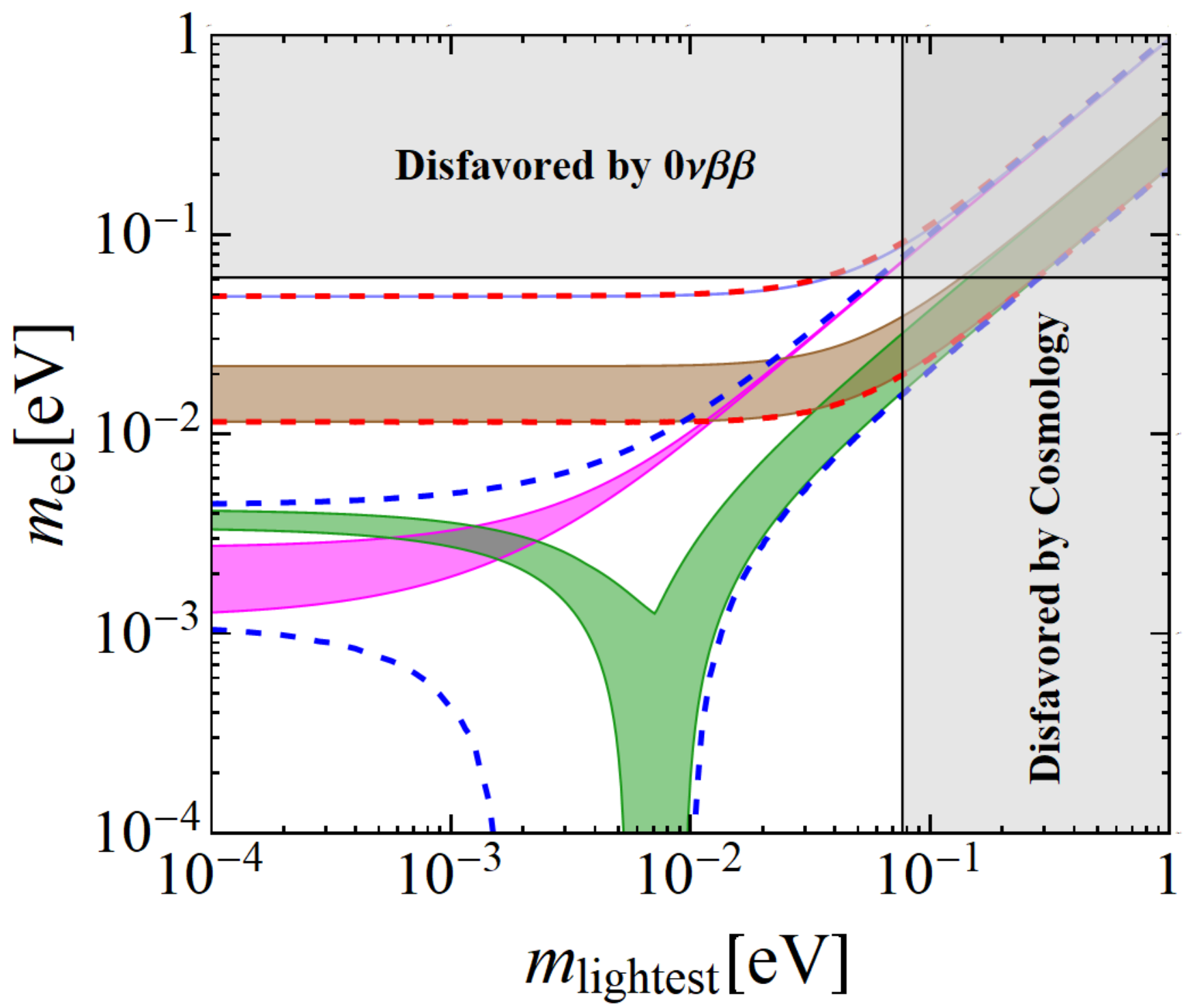}\\[0.05in]
\hskip-0.1in\includegraphics[width=0.90\linewidth]{legend_mee_rasterized.pdf}
\end{tabular}
\caption{\label{mee-e-tau}The same as figure~\ref{mee-e-mu} for the
  generalized $e-\tau$ reflection. The CP label $\Theta$ is
  free on the left panel, and we fix $\Theta=\pi/9$ and
  $\Theta=8\pi/9$ on the right panel. Notice that the IO: $0\leq \Theta \leq \frac{\pi}{2}$ region (cyan) is very narrow and is  close to the maximum $m_{ee}$ for IO. This region can be tested very soon in neutrinoless double beta decay experiments. }
\end{center}
\end{figure}

\begin{figure}[hptb!]
\begin{center}
\begin{tabular}{c} \hskip-0.6in
\includegraphics[width=0.5\linewidth]{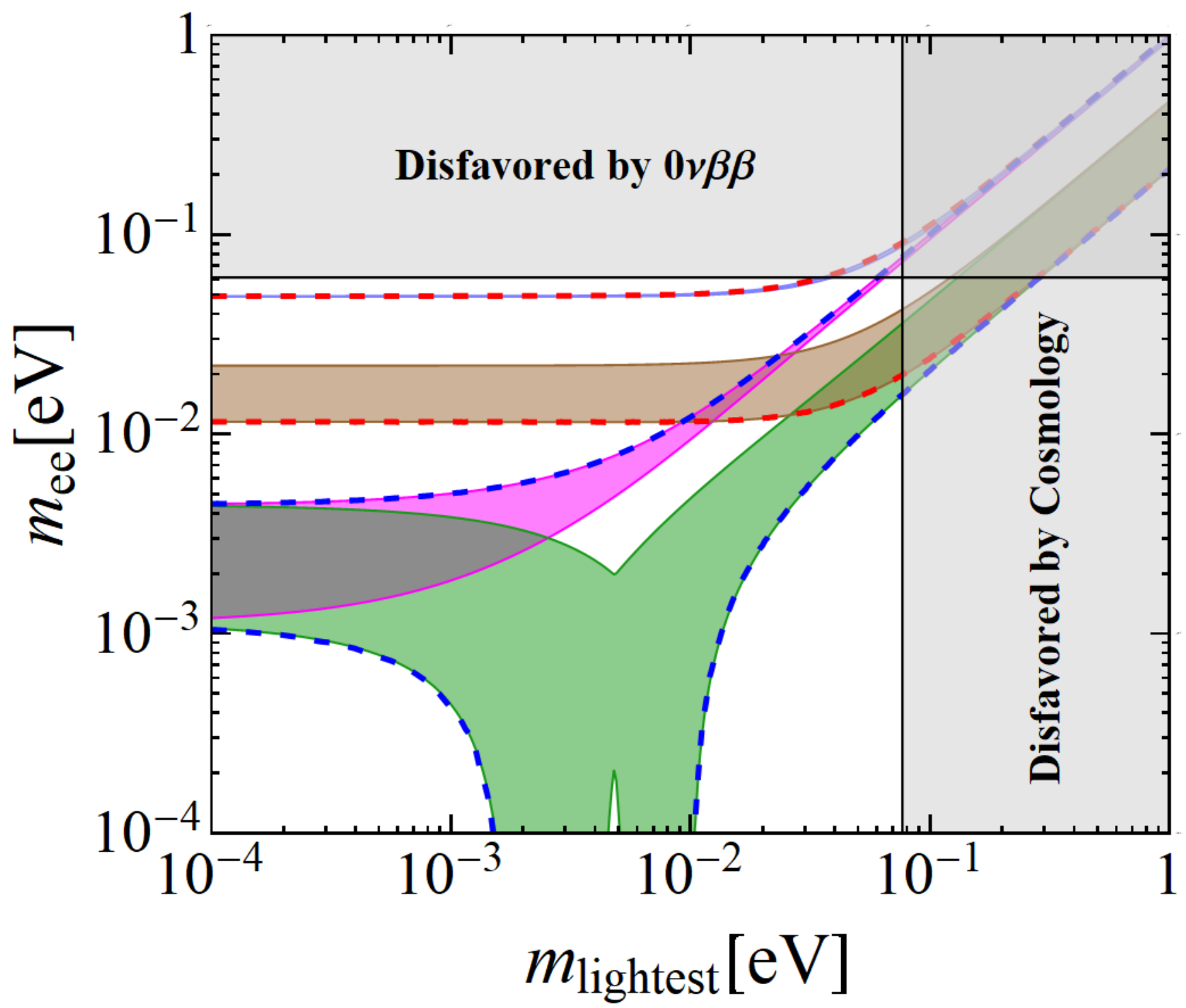}
\includegraphics[width=0.5\linewidth]{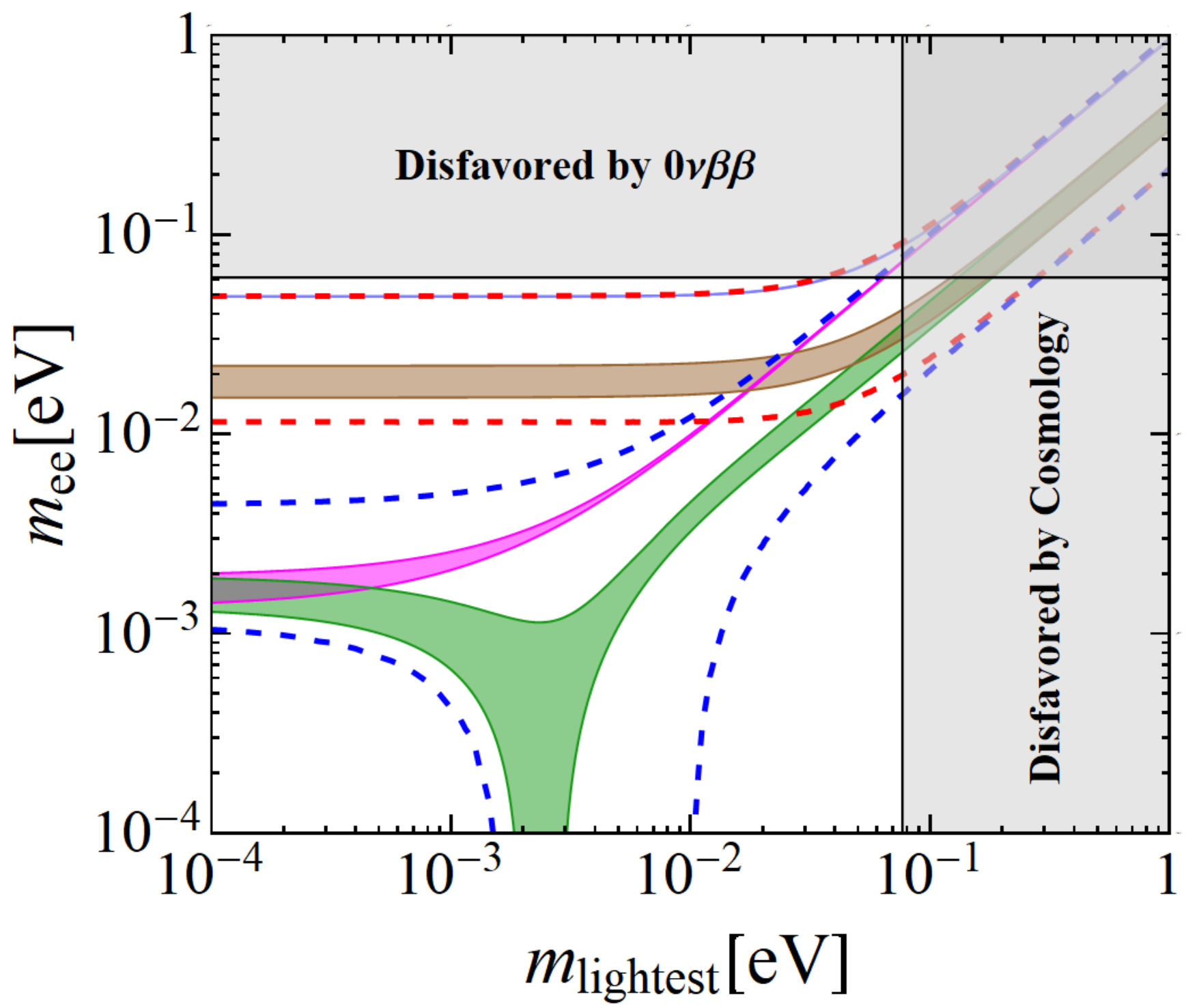}\\[0.05in]
\hskip-0.1in\includegraphics[width=0.90\linewidth]{legend_mee_rasterized.pdf}
\end{tabular}
\caption{\label{mee-mu-tau}The same as figure~\ref{mee-e-mu} for the
generalized $\mu-\tau$ reflection. The CP label $\Theta$ is free
on the left panel, and we fix $\Theta=\pi/6$ and $\Theta=5\pi/6$ on
the right panel.  Notice that the IO: $0\leq \Theta \leq \frac{\pi}{2}$ region (cyan) is very narrow and is  close to the maximum $m_{ee}$ for IO. This region can be tested very soon in neutrinoless double beta decay experiments. }
\end{center}
\end{figure}

Similar neutrinoless double beta decay predictions can also be
obtained for the cases of $e-\tau$ and $\mu-\tau$ symmetries, as shown
in Fig.~\ref{mee-e-tau} and Fig.~\ref{mee-mu-tau} respectively.
One sees that the generalized CP symmetry allows only a restricted
range for $m_{ee}$, and thus can be used to test the predictions of
the generalized CP symmetries. In particular, if we fix the CP
label $\Theta$, then the allowed range for $m_{ee}$ becomes much
narrower for both mass orderings, and also for both cases of $e-\tau$
and $\mu-\tau$ symmetries. Future neutrinoless double decay
experiments will probe almost all of the IO region, thus allowing to
distinguish between $0\leq\Theta<\pi/2$ and $\pi/2\leq\Theta\leq\pi$.


%%%%%%%%%%%%%%%%%%%%%%%%%%%%%%%%%%%%%%%%%%%%%%%%%%%%%%%%%%%%%%%%%%%%%%%%%%%%%%%

\section{\label{sec:conclusion}Conclusion}

%%%%%%%%%%%%%%%%%%%%%%%%%%%%%%%%%%%%%%%%%%%%%%%%%%%%%%%%%%%%%%%%%%%%%%%%%%%%%%%

The imposition of generalized CP symmetries provides a powerful framework for predicting the lepton mixing angles and
phases.  We have investigated the theory of generalized CP
transformations acting on the mass matrices of charged leptons.
We have considered in detail the case of generalized $e-\mu$, $e-\tau$
and $\mu-\tau$ symmetries.
The basic tool is the Takagi factorization which is used to
express the physical parameters of the lepton mixing matrix, three
mixing angles, and three CP phases, in terms of a restricted set of
independent ``theory parameters'' (labels) associated with a
given choice of the CP transformation.
Current best fit values of the mixing angles and the favored value for
$\delta_{CP}=3\pi/2$ constrain the allowed ``theory'' values of the labels characterizing the CP transformation, for example
$\Theta$.
In each case we have obtained strong correlations involving the mixing
angles and CP phases, valid both for Majorana and Dirac neutrinos see,
for example, in Eqs.~(\ref{eq:t12_dcp}, \ref{eq:e-tau_corr1},
\ref{eq:corr1_mu-tau}). Specific benchmark model examples were given
in the tables.
Our predictions for the leptonic CP violating phase $\delta_{CP}$,
summarized in Fig.~\ref{delta_cp}, provides model--independent probes
of our underlying generalized CP symmetry approach at upcoming long
baseline oscillation experiments, such as DUNE, aimed at the
measurement of CP violation.
For the case of Majorana neutrinos, our predictions for the neutrino
mixing angles and phases also include the two Majorana phases, as seen
in Figs.~\ref{mee-the-phi} and \ref{mee-phi-phi}.
We have also derived the resulting predictions for the effective mass
parameter $m_{ee}$ characterizing the neutrinoless double beta decay
rates.
Predicted ranges for $m_{ee}$ in each case can be used to test the
residual CP symmetry hypothesis at the uponming generation of
sensitive experiments, such as KamLAND-ZEN, CUORE, LEGEND, nEXO and
NEXT. We would like to remind the readers that the effect of the CP labels $\alpha$, $\beta$ and $\gamma$ is to shift the Majorana phases $\phi_{12}$ and $\phi_{13}$. The numerical results in Figs. \ref{mee-e-mu}, \ref{mee-e-tau} and \ref{mee-mu-tau}
%for $\phi_{12}$, $\phi_{13}$ and the neutrinoless double beta decay amplitude parameter
for $m_{ee}$
% in this paper
are obtained under the assumption of $\alpha=\beta=\gamma=0$.

Notice that, although we have treated in full generality the
implications of generalized CP symmetries, we have not attempted to
obtain a rationale for their possible origin. In fact this is an
interesting open issue that deserves a dedicated study, which goes far
beyond the scope of this present paper and that will be tackled
elsewhere.

%%%%%%%%%%%%%%%%%%%%%%%%%%%%%%%%%%%%%%%%%%%%%%%%%%%%%%%%%%%%%%%%%%%%%%%%%%%
\begin{acknowledgments}

  Work supported by the Spanish grants FPA2017-85216-P and
  SEV-2014-0398 (MINECO), PROMETEOII/2014/084 (Generalitat
  Valenciana), and by the National Natural Science Foundation of
  China, Grant No 11522546. P.C. and G.J.D acknowledge J.N. Lu for his
  kind help on plotting the figures.

\end{acknowledgments}

%%%%%%%%%%%%%%%%%%%%%%%%%%%%%%%%%%%%%%%%%%%%%%%%%%%%%%%%%%%%%%%%%%%%%%%%%%

%\bibliographystyle{bib_style_T1}
%\bibliography{local/all,local/valle}

%\bibliographystyle{bib_style_T1}
%\bibliographystyle{utphys}
%\bibliography{gcp}


\end{document}